\documentclass[aps,pra,twocolumn,showpacs,superscriptaddress,floatfix]{revtex4-1}
\usepackage{amsmath}
\usepackage{graphics}
\usepackage{graphicx}
\usepackage{color}
\begin{document}

\title{The prospects of forming ultracold molecules in $^2\Sigma$ states \\
by magnetoassociation of alkali-metal atoms with Yb}
\author{Daniel A. Brue}
\author{Jeremy M. Hutson}
\email{J.M.Hutson@durham.ac.uk} \affiliation{Joint Quantum Centre (JQC)
Durham/Newcastle, Department of Chemistry, Durham University, South Road,
Durham, DH1 3LE, United Kingdom}
\date{\today}
\begin{abstract}
We explore the feasibility of producing ultracold diatomic molecules with
nonzero electric and magnetic dipole moments by magnetically associating two
atoms, one with zero electron spin and one with nonzero spin. Fesh\-bach
resonances arise through the dependence of the hyperfine coupling on
internuclear distance. We survey the Feshbach resonances in diatomic systems
combining the nine stable alkali-metal isotopes with those of Yb, focussing on
the illustrative examples of RbYb and CsYb. We show that the resonance widths
may expressed as a product of physically comprehensible terms in the framework
of Fermi's Golden Rule. The resonance widths depend strongly on the background
scattering length, which may be adjusted by selecting the Yb isotope, and on
the hyperfine coupling constant and the magnetic field. In favorable cases the
resonances may be over 100 mG wide.
\end{abstract}
\maketitle

\section{Introduction}
\label{sec:intro} The successes of cooling gases of atoms to ultracold
temperatures have led to great interest in producing molecules at similar
temperatures. Because molecules have a richer internal structure and more
complex interactions than atoms, ultracold ($\mu$K) molecules offer the
possibility of exploring a wide range of new research areas, including
high-precision measurement \cite{PhysRevLett.96.143004, PhysRevLett.92.133007,
PhysRevLett.89.023003}, quantum information \cite{PhysRevLett.88.067901,
PhysRevLett.82.1975} and quantum simulation \cite{Micheli:2006}.

Molecules may be formed in ultracold atomic gases either by photoassociation
\cite{RevModPhys.78.483} or by magnetoassociation \cite{RevModPhys.78.1311}. In
the latter, cold atomic clouds are subjected to time-dependent magnetic fields
that convert atom pairs into molecules by adiabatic passage across zero-energy
Feshbach resonances \cite{RevModPhys.82.1225}. Recent years have seen
substantial progress in producing ultracold molecules made up of pairs of
alkali-metal atoms \cite{Herbig:MolGas, Regal:MolGas, Strecker:MolGas,
Cubizolles:MolGas, Grimm:Li2BEC, Ni:KRb, Danzl:Cs2, Lang:Rb2, Heo:2012}. The
molecules are left in high vibrational states and are susceptible to
collisional trap loss. For KRb \cite{Ni:KRb}, Cs$_2$ \cite{Danzl:Cs2}, and
triplet Rb$_2$ \cite{Lang:Rb2}, it has been possible to transfer the molecules
to the absolute ground state by Stimulated Raman Adiabatic Passage (STIRAP).

There is now great interest in the formation of cold molecules that have both
electric and magnetic dipole moments \cite{PSZ:RbSr, Gupta:LiYb,
Gupta:PRA:LiYb, Takahashi:LiYb, Gorlitz:RbYb, Brue:LiYb:2012}. Such molecules
offer additional possibilities for manipulation, trapping, and control because
they can be influenced by both electric and magnetic fields. In the present
paper we investigate the prospects for magnetoassociation of alkali-metal atoms
(Alk) with $^1$S atoms (specifically Yb) to form $^2\Sigma$ heteronuclear
diatoms with electron spin $S=1/2$.

Ytterbium is an excellent candidate for pairing with the alkali metals. It has
7 stable isotopes (5 zero-spin bosons, 2 fermions), and a closed-shell,
singlet-spin electronic structure. Both bosonic \cite{Takasu:2003,
Fukuhara:boson:2007, Fukuhara:2009, Sugawa:2011} and fermionic
\cite{Fukuhara:Yb-173:2007, Fukuhara:fermion:2007} isotopes have been cooled to
quantum degeneracy.
%For Hg, both bosonic and fermionic isotopes have been trapped
%\cite{Hachisu:2008} and fermionic isotopes have been cooled to sub-Doppler
%temperatures \cite{McFerran:2010}. In the present paper, we principally
%consider Alk-Yb systems, but also consider the possibilities for Alk-Hg
%systems.
Different isotopic combinations have different scattering lengths, and produce
molecules with different binding energies; they thus have Feshbach resonances
at different magnetic fields.

The existence of magnetically tunable Feshbach resonances requires coupling
between a continuum scattering state of the atomic pair and a molecular state
that crosses it as a function of magnetic field. For pairs of alkali-metal
atoms, this coupling is provided by the difference between the singlet and
triplet potential curves and by the magnetic dipolar interaction between the
electron spins. However, neither of these effects exists in systems of the type
considered here. Instead, the most significant coupling between the atomic and
molecular states is provided by the $R$-dependence in the hyperfine coupling
constant of the alkali-metal atom \cite{PSZ:RbSr}. Such $R$-dependences exist
in alkali dimers \cite{Strauss:2010}, but in that case they merely produce
small shifts in bound state energies and resonance positions, rather than
driving new resonances. If the closed-shell atom has non-zero nuclear spin, it
can also couple to the unpaired electron spin. For the case of LiYb
\cite{Brue:LiYb:2012}, this coupling has been found to be much stronger than
that due to the Li nucleus. However, this latter effect is less important for
the heavier alkali-metal atoms considered here, where the coupling to the
alkali-metal nucleus itself is stronger.

In previous work, we extracted resonance positions and widths for RbSr
\cite{PSZ:RbSr} and LiYb \cite{Brue:LiYb:2012} from coupled-channel quantum
scattering calculations. In the present paper, we extend these studies to a
range of heavier systems and show how the widths may be broken down into their
contributing factors within the framework of Fermi's Golden Rule.

The theoretical development presented here is applicable to any system made up
of an alkali-metal atom paired with a closed-shell atom. In the present study,
we have considered the whole range of Alk-Yb systems, but we focus our
presentation on the illustrative examples of Rb-Yb, for which the scattering
lengths are approximately known, and Cs-Yb, for which they are as yet unknown.
In section \ref{sec:theory} we describe the theoretical methods used. In
section \ref{sec:results} we present our results, with discussion of system
characteristics that lead to Feshbach resonances suitable for molecule
formation.

\section{Theory}
\label{sec:theory}

%%%%%%%%%%%%%%%%%%%%%%%%%%%%%%%%%%%%%%%%%%%%%%%%%%%%%%%%%%%%%%%%%%%%%%%%%%%%%%%%%%%%%%%%%%
\subsection{Collisions between alkali-metal and closed-shell atoms}
The Hamiltonian for an alkali-metal atom $a$ in a $^2$S state, interacting with
a closed-shell atom $b$ in a $^1$S state, is
\begin{equation}
\hat{H}=\frac{\hbar^2}{2\mu}\left[-\frac{{\rm d^2}}{{\rm d}R^2}
+\frac{\hat{L}^2}{R^2}\right]+\hat{U}(R)+\hat{H}_a + \hat{H}_b
\label{eqn:Ham}
\end{equation}
where $\hat{L}$ is the two-atom rotational angular momentum operator and
$\hat{U}(R)$ is the interaction operator. $\hat{H}_a$ and $\hat{H}_b$ are the
single-atom hamiltonians,
\begin{eqnarray}
\label{eqn:BRHama}
H_a&=&\zeta_a\hat i_a\cdot\hat s + \left(g_a\mu_{\rm N}\hat i_{a,z}
+ g_{e}\mu_{\rm B}\hat s_z\right)B \\
H_b&=&g_b\mu_{\rm N}\hat i_{b,z}B,
\label{eqn:BRHamb}
\end{eqnarray}
where $\hat s$, $\hat i_a$ and $\hat i_b$ are the electron and nuclear spin
operators, $g_e$, $g_a$ and $g_b$ are the electronic and nuclear $g$ factors,
and $\mu_{\rm B}$ and $\mu_{\rm N}$ are the Bohr and nuclear magnetons.
$\zeta_a$ is the hyperfine coupling constant for the alkali-metal atom and $B$
is the external magnetic field, whose direction defines the $z$ axis. In the
present work we use lower-case angular momentum operators and quantum numbers
for individual atoms and upper-case for the corresponding molecular quantities.

The interaction of a $^2$S atom with a $^1$S atom produces only one molecular
electronic state, of $^2\Sigma$ symmetry. However, the hyperfine coupling
constant of the alkali-metal atom is modified by the presence of the
closed-shell atom \cite{PSZ:RbSr}, and if $i_b\ne0$ then there is also
hyperfine coupling involving the nucleus of atom $b$ \cite{Brue:LiYb:2012},
\begin{eqnarray}
\zeta_a(R) &=& \zeta_a + \Delta\zeta_a(R); \\
\zeta_b(R) &=& \Delta\zeta_b(R). \label{eqn:zetaR}
\end{eqnarray}
The interaction operator $\hat U(R)$ is thus
\begin{equation}
\hat U(R) = V(R) + \Delta\zeta_a(R) \hat i_a\cdot \hat s
+ \Delta\zeta_b(R) \hat i_b\cdot \hat s,
\label{eqn:vhat}
\end{equation}
where $V(R)$ is the electronic interaction potential. Most of the theory
presented here remains applicable when atom $a$ is a non-alkali-metal atom in a
multiplet-S state.

\begin{figure}
\includegraphics[width=\linewidth]{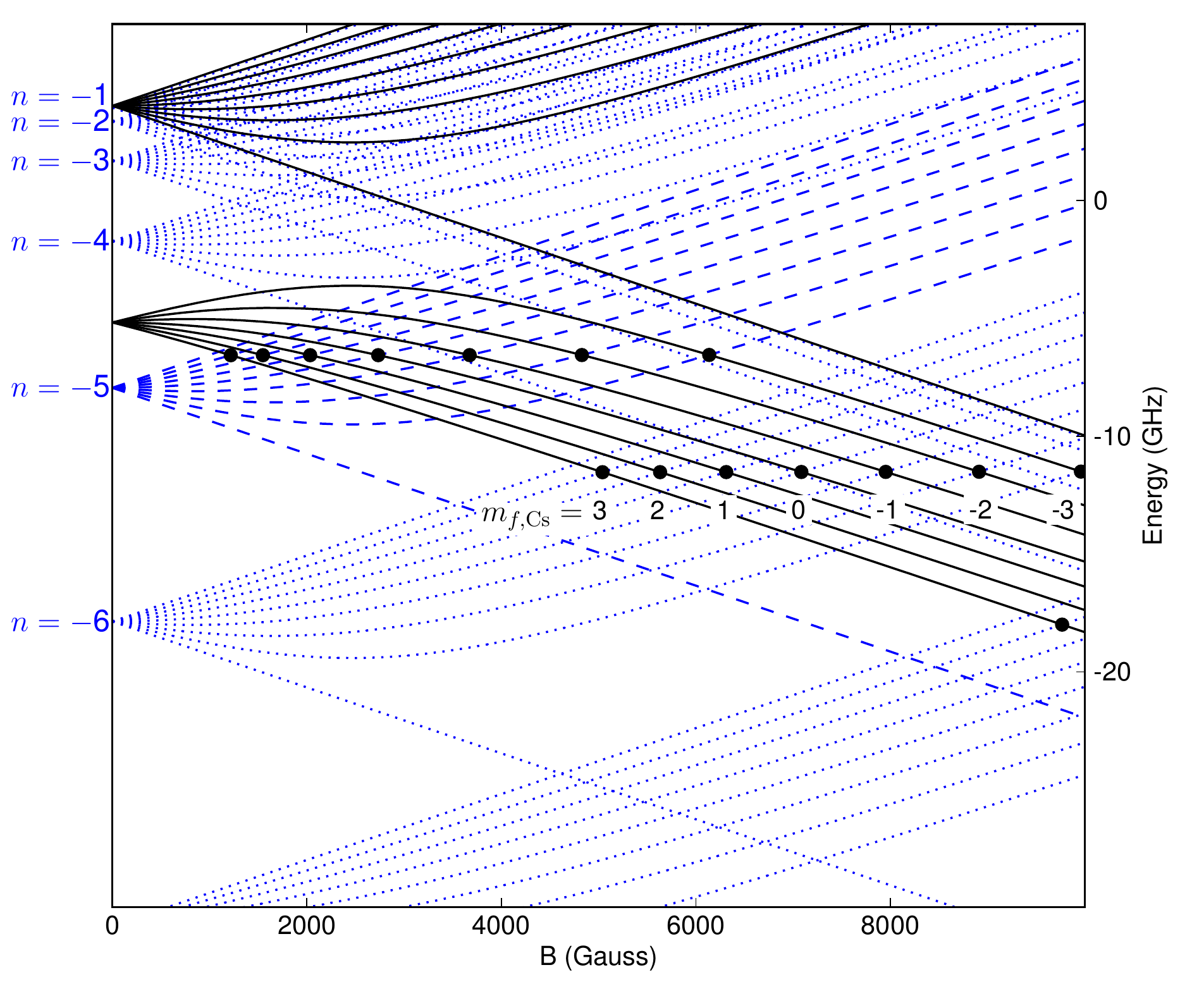}
\caption{(color online). Hyperfine energy levels for $^{133}$Cs in its ground
$^2$S state (solid black lines) and for near-threshold states of CsYb arising
from the upper hyperfine manifold $|\alpha_2,m_{f,{\rm Cs}}\rangle$ (dashed
blue lines), calculated for a potential with a scattering length of $-38$ bohr.
The $n=-1$ level is almost hidden by the threshold. The solid circles mark the
Feshbach resonances that occur at crossings between bound states and atomic
thresholds with the same value of $m_{f,{\rm Cs}}$. } \label{fig:BreitRabi1}
\end{figure}

Figure \ref{fig:BreitRabi1} shows the energy levels of the $^{133}$Cs atom,
with $i_a=7/2$ (black solid lines). At zero field the levels may be labeled by
quantum numbers $f_a,m_{f,a}$, where $f_a=i_a\pm 1/2$, whereas at high field
the nearly good quantum numbers are $m_{s,a}$ and $m_{i,a}$. In the present
paper we indicate the lower and upper states for each $m_{f,a}=m_{s,a}+m_{i,a}$
as $|\alpha_1,m_{f,a}\rangle$ and $|\alpha_2,m_{f,a}\rangle$ respectively.

The Hamiltonian (\ref{eqn:Ham}) may be written as the sum of a zeroth-order
term $\hat H^0$ and a perturbation $\hat H'$,
\begin{eqnarray}
\hat{H}^0&=&\frac{\hbar^2}{2\mu}\left[-\frac{{\rm d}^2}
{{\rm d}R^2}+\frac{\hat{L}^2}{R^2}\right]+V(R)+\hat{H}_a + \hat{H}_b; \\
\label{eqn:H0}
\hat{H}'&=&\Delta\zeta_a(R) \hat i_a\cdot \hat s
+ \Delta\zeta_b(R) \hat i_b\cdot \hat s.
\label{eqn:Hprime}
\end{eqnarray}
The zeroth-order Hamiltonian is separable, and its eigenfunctions are products
of atomic functions $|\alpha_i,m_{f,a}\rangle|i_b,m_{i,b}\rangle$ and radial
functions $\psi(R)$. The latter are eigenfunctions of the 1-dimensional
Hamiltonian
\begin{equation}
\frac{\hbar^2}{2\mu}\left[-\frac{{\rm d}^2} {{\rm
d}R^2}+\frac{L(L+1)}{R^2}\right] + V(R), \label{eqn:Hrad}
\end{equation}
with eigenvalues $E_n$. The eigenvalues of $\hat H^0$ are $E_n + E_a + E_b$,
where $E_a$ and $E_b$ are the eigenvalues of $\hat H_a$ and $\hat H_b$. By
contrast with the alkali-metal dimers, the molecular states thus lie almost
parallel to the atomic states as a function of magnetic field. They also have
almost exactly the same spin character. The only terms in the Hamiltonian
(\ref{eqn:Ham}) that couple $|\alpha_1,m_{f,a}\rangle|i_b,m_{i,b}\rangle$ and
$|\alpha_2,m_{f,a}'|i_b,m_{i,b}'\rangle\rangle$ are the weak couplings
involving $\Delta\zeta_a(R)$ and $\Delta\zeta_b(R)$. The former couples states
with $m_{f,a}'=m_{f,a}$ and $m_{i,b}'=m_{i,b}$, while the latter couples states
with $m_{f,a}'=m_{f,a}\pm1$ and $m_{i,b}'=m_{i,b}\mp1$.

The Hamiltonian (\ref{eqn:Ham}) is entirely diagonal in $L$, so resonances in
s-wave scattering can be caused only by $L=0$ bound states. The only
interactions off-diagonal in $L$ are spin-rotation and nuclear quadrupole
interactions, which are neglected in the present work. This again contrasts
with the alkali-metal dimers, where the magnetic dipolar interaction between
the electron spins and second-order spin-orbit coupling provide relatively
strong interactions that produce resonances from bound states with $L>0$ in
s-wave scattering.

Figure \ref{fig:BreitRabi1} shows the highest few vibrational states for CsYb
with spin character $|\alpha_2,m_{f,a}'\rangle$ and vibrational quantum numbers
$n=-1$, $-2\ldots -7$ (with respect to threshold) as dashed blue lines,
calculated for a potential $V(R)$ with an s-wave scattering length $a=-38$
bohr. The couplings involving $\Delta\zeta_a(R)$ give rise to Feshbach
resonances at fields where bound states $|\alpha_2,m_{f,a},n\rangle$ cross
thresholds $|\alpha_1,m_{f,a}\rangle$, shown as solid circles in Figure
\ref{fig:BreitRabi1}. In the present work we neglect couplings due to
$\Delta\zeta_b(R)$ and set $i_b=0$ for all isotopes. This will give accurate
results for resonances with $m_{f,a}'=m_{f,a}$ but will suppress resonances
with $m_{f,a}'=m_{f,a}\pm1$, which actually exist for $^{171}$Yb and
$^{173}$Yb.

%%%%%%%%%%%%%%%%%%%%%%%%%%%%%%%%%%%%%%%%%%%%%%%%%%%%%%%%%%%%%%%%%%%%%%%%%%%%%%%%%%%%%%%%%%
\subsection{Electronic Structure Calculations}

\subsubsection{Potential Energy Curves}

We have constructed the electronic potential energy curves $V(R)$ by carrying
out electronic structure calculations at short and medium range and switching
to a form incorporating dispersion interactions at long range.

\begin{figure}
\includegraphics[width=\linewidth]{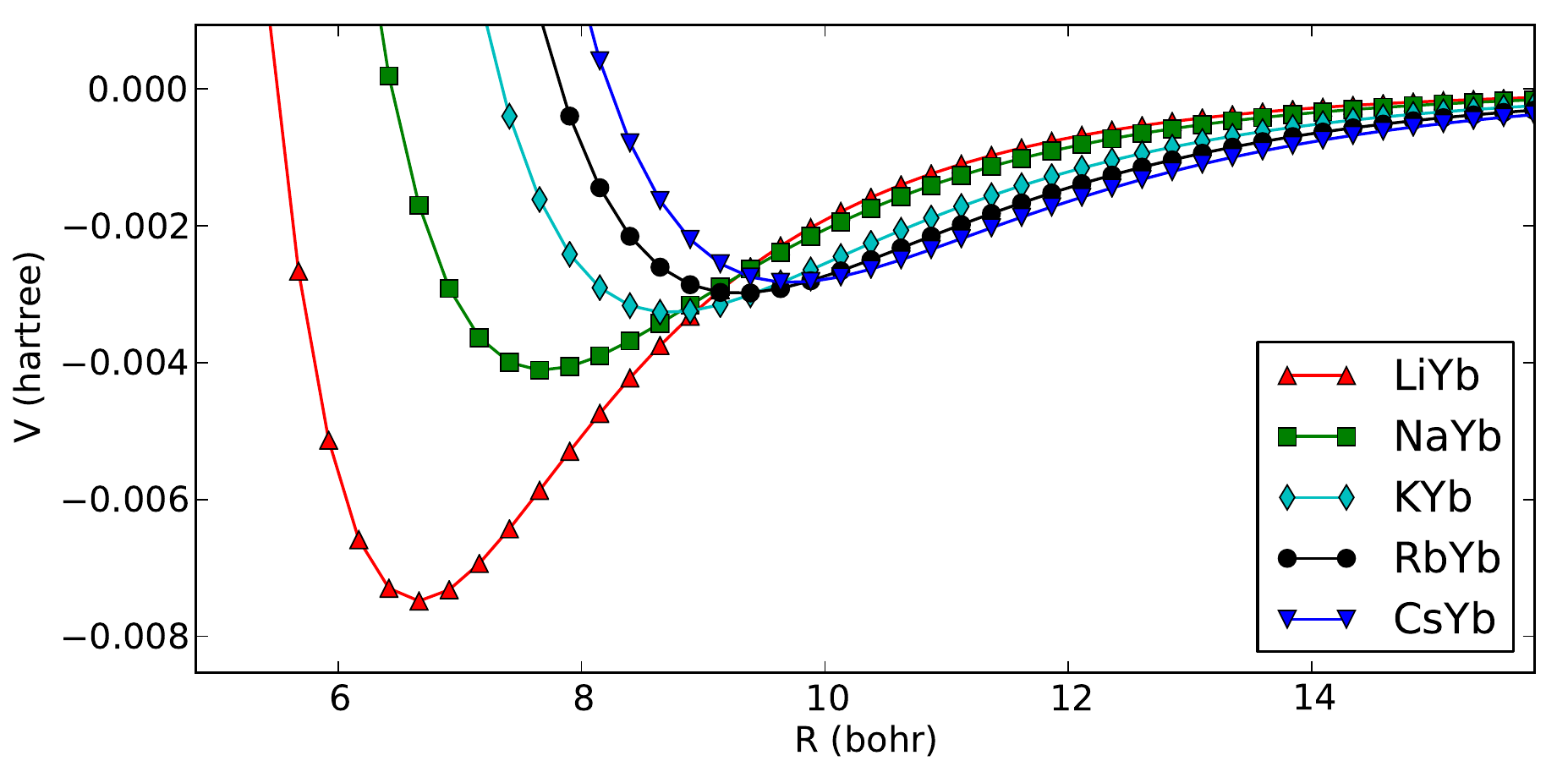}
\caption{(color online). Electronic potential energy curves $V(R)$ from CCSD(T)
calculations on the Alk-Yb systems.}\label{fig:AlkYbPots}
\end{figure}

We obtained ground-state potential curves for NaYb, KYb, RbYb and CsYb from
CCSD(T) calculations (coupled-cluster with single, double, and non-iterative
triple excitations) using the \textsc{Molpro} package \cite{MOLPRO_brief:2006}.
For Yb, we used the quasi-relativistic effective core potential (ECP) of Dolg
\textit{et al.}\ and its corresponding basis set \cite{DolgYbMWB}, with 60
electrons in the inner 4 shells represented by the ECP and the remaining 10
electrons ($p^6s^4$) treated explicitly.
%For Hg we used the quasi-relativistic ECP of Andrae \textit{et al.}\
%\cite{Andrae:1990} for the inner 60 electrons, and the corresponding published
%basis for the outer 20 electrons.
ECPs \cite{Leininger1996274} and their corresponding basis sets
\cite{StuttgartECPs} were also used for K, Rb, and Cs. An ECP was not used for
Na; all 11 electrons were represented with the cc-pvqz basis set of Prascher
\textit{et al.}\ \cite{NaBasis}. For each system, CCSD(T) calculations were
carried out at a series of points from 2 to 40 bohr and the potentials were
then interpolated using the reproducing kernel Hilbert space (RKHS) method
\cite{RKHS}. The resulting potential curves are shown in Figure
\ref{fig:AlkYbPots}, together with the LiYb curve of Zhang {\em et al.}\
\cite{Zhang:2010}, obtained using similar methods but with a fully relativistic
ECP for Yb. The well depths and equilibrium distances are given in Table
\ref{tbl:PotData}.

At long range the potential curves were represented as
\begin{equation}
V(R) = -C_6 R^{-6} -C_8 R^{-6} -C_{10} R^{-10}.
\label{eqn:long}
\end{equation}
The $C_6$ coefficients used for the long-range potential were obtained from
Tang's combination rule \cite{PhysRev.177.108} based on the Slater-Kirkwood
formula,
\begin{equation}
C_6^{ab} = \frac{C_6^{aa}C_6^{bb}\alpha^a(0)\alpha^b(0)}
{C_6^{aa}(\alpha^b(0))^2+C_6^{bb}(\alpha^a(0))^2}
\label{eqn:TangSK}
\end{equation}
using the homonuclear $C_6$ coefficients for Alk-Alk \cite{Derevianko2010323}
and Yb-Yb \cite{PhysRevA.77.012719} and the static polarizabilities $\alpha(0)$
for the alkali-metal atoms \cite{Derevianko2010323} and Yb \cite{YbAlpha0}.
Equation (\ref{eqn:TangSK}) gives $C_6$ coefficients well within 1\% of the
values of ref.\ \cite{Derevianko2010323} for all the mixed alkali-metal pairs.
The results for the Alk-Yb systems are included in Table \ref{tbl:PotData}. The
$C_8$ and $C_{10}$ terms were omitted except when fitting to the experimental
spectra for RbYb as described in section \ref{sec:results:rbyb} below. The
short-range and long-range regions of the potential were joined using the
switching function of Janssen {\em et al.}\ \cite{Janssen:2009} between the
distances 28 and 38 bohr.

\begin{table}
\begin{center}
\begin{tabular}{rrrr}
\hline\hline
System & $R_e$ & $V(R_e)$ & $C_6$ \\
       & (bohr)& (${\rm mE}_{\rm h}$) & (${\rm E}_{\rm h}a_0^6$) \\
\hline
LiYb & 6.65 & $-7.48$ & 1594 \\ %check that Re is the Zhang value
NaYb & 7.61 & $-4.61$ & 1690 \\
 KYb & 8.88 & $-3.36$ & 2580 \\
RbYb & 9.28 & $-2.99$ & 2830 \\
CsYb & 9.72 & $-2.83$ & 3370 \\
\hline\hline
\end{tabular}
\end{center}
\caption{Properties of the interaction potentials used in the present work. The
well depths and equilibrium distances are from CCSD(T) calculations and the
$C_6$ coefficients are from Eq.\ (\ref{eqn:TangSK}).} \label{tbl:PotData}
\end{table}

\subsubsection{Hyperfine Coupling}
The hyperfine coupling constant of an atom is a measure of the interaction
between its nuclear spin and the electron spin density at the nucleus, which in
the case of an alkali metal comes principally from the single valence electron.
Approach of another atom perturbs the electronic wavefunction and alters the
spin density at the nucleus, so that the coupling between the electron and
nuclear spins becomes a function of internuclear distance $R$.

\begin{figure}
\includegraphics[width=\linewidth]{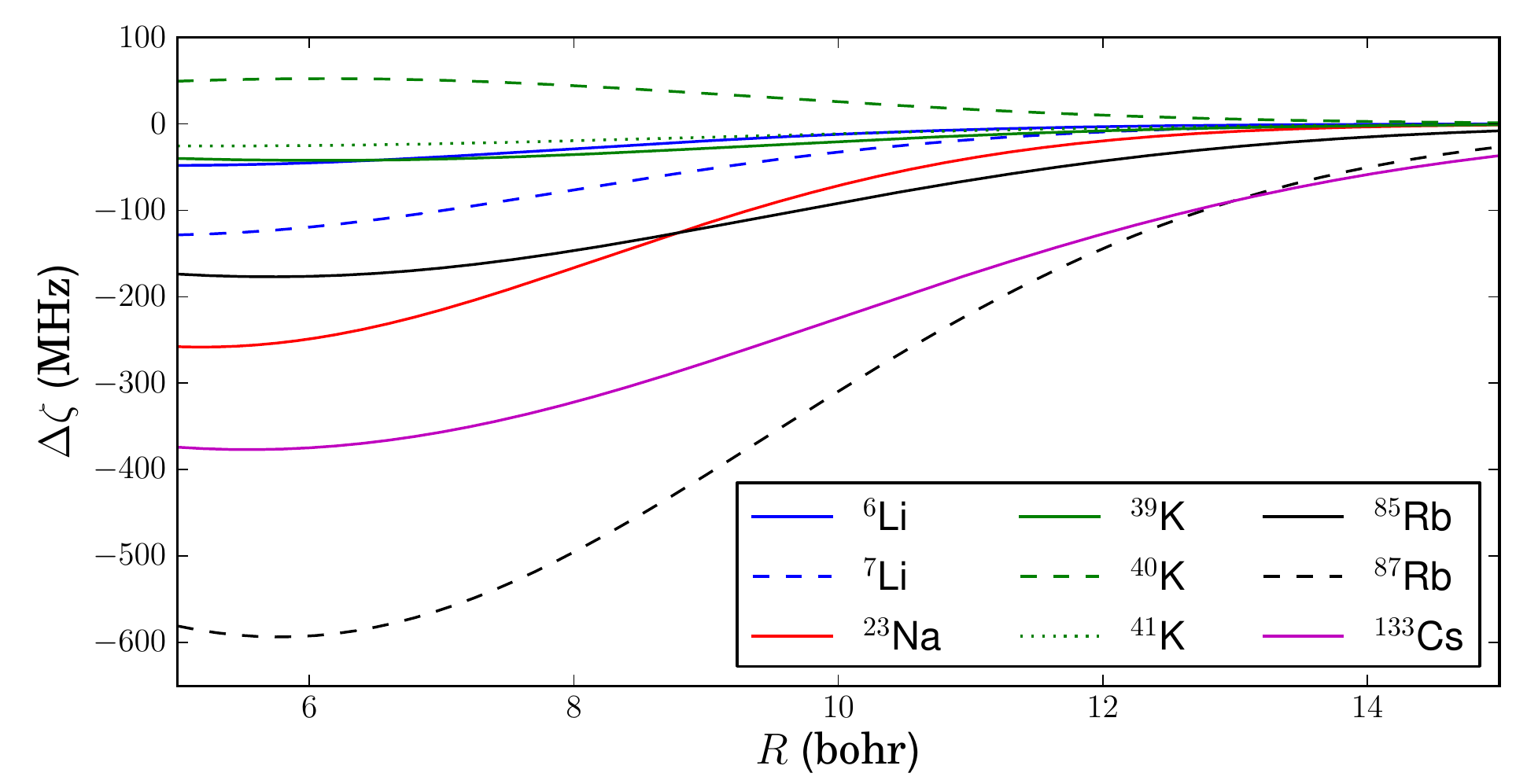}
\caption{(color online). Distance dependence $\Delta\zeta_a(R)$ of the
hyperfine coupling constants for the Alk-Yb systems.}
\label{fig:dzetas}
\end{figure}

We have calculated the hyperfine coupling constants $\zeta_a(R)$ for the Alk-Yb
systems, using density-functional theory with the KT2 functional
\cite{Keal:2003}, as implemented in the ADF suite of programs \cite{ADF}. We
fitted these results to a variety of functional forms and found that, in the
range of $R$ for which the vibrational wavefunctions are non-zero, a Gaussian
function $\Delta\zeta_a(R) = \zeta_0 e^{-\beta(R-R_c)^2}$ gave an adequate fit
to the DFT results. These functions are shown in Figure \ref{fig:dzetas} for
each of the Alk-Yb systems and the parameters are given in Table
\ref{tbl:dzetas}.

\begin{table}
\begin{center}
\begin{tabular}{rrrr}
\hline\hline
 &        $\zeta_0$ (MHz) & $\beta$ (bohr$^{-2})$ & $R_c$ (bohr) \\
\hline
%$^6$Li    & $-48.2$  &  0.0535 &  4.92 \\
$^6$Li    & $-48.8$  &  0.0535 &  4.92 \\
%$^7$Li    & $-129.$  &  0.0503 &  4.78 \\
$^7$Li    & $-129.$  &  0.0535\footnote{The value of
$\beta$ for Li was reported incorrectly in ref.\ \cite{Brue:LiYb:2012}.} &  4.92 \\
$^{23}$Na & $-258.$  &  0.0553 &  5.18 \\
$^{39}$K  & $-42.4$  &  0.0474 &  6.09 \\
$^{40}$K  & $52.5$   &  0.0474 &  6.09 \\
$^{41}$K  & $-23.3$  &  0.0474 &  6.09 \\
%$^{40}$K  & $52.3$   &  0.0479 &  6.11 \\
%$^{41}$K  & $-25.6$  &  0.0367 &  5.30 \\
$^{85}$Rb & $-177.$  &  0.0357 &  5.71 \\
$^{87}$Rb & $-597.$  &  0.0357 &  5.71 \\
%$^{87}$Rb & $-593.$  &  0.0364 &  5.77 \\
$^{133}$Cs & $-377.$  &  0.0260 &  5.54 \\
\hline\hline
\end{tabular}
\end{center}
\caption{Parameters of the Gaussian functions used to represent
$\Delta\zeta_a(R)$, the distance-dependence of the hyperfine coupling constant
of an alkali-metal atom interacting with Yb.} \label{tbl:dzetas}
\end{table}

%%%%%%%%%%%%%%%%%%%%%%%%%%%%%%%%%%%%%%%%%%%%%%%%%%%%%%%%%%%%%%%%%%%%%%%%%%%%%%%%%%%%%%%%%%
\subsection{Resonance widths from coupled-channel calculations}

Near resonance, the s-wave scattering length $a(B)$ as a function of magnetic
field $B$ behaves as \cite{PhysRevA.51.4852}
\begin{equation}
a(B)=a_{\rm bg}\left(1-\frac{\Delta}{B-B_{\rm res}}\right),
\label{eqn:awrtB}
\end{equation}
where $B_{\rm res}$ is the resonance position and $a_{\rm bg}$ is the
background scattering length. The magnitude of the resonance width, $\Delta$,
is critical for determining whether magnetoassociation is experimentally
feasible. Defining $B_{\rm zero}$ as the field where $a(B)=0$ near resonance,
Eq.\ (\ref{eqn:awrtB}) implies $\Delta = B_{\rm zero} - B_{\rm res}$.

In the present work, we obtained scattering lengths $a(B)$ principally from
coupled-channel calculations. The coupled equations for each field $B$ were
constructed in an uncoupled basis set $|s_a m_{s,a}\rangle |i_a m_{i,a}\rangle
|L M_L\rangle$ and solved using the MOLSCAT package \cite{molscat:v14-short,
Gonzalez-Martinez:2007}. The s-wave scattering length was then obtained from
the identity \cite{Hutson:res:2007} $a=(ik)^{-1}(1-S_{00})/(1+S_{00})$, where
$S_{00}$ is the diagonal S-matrix element in the incoming channel,
$k=\hbar^{-1}(2\mu E_{\rm col})^{1/2}$, and the collision energy $E_{\rm col}$
was taken to be 1~nK~$\times{k_{\rm B}}$. MOLSCAT has an option to converge
numerically on the fields corresponding to both poles and zeroes in $a(B)$,
allowing the extraction of $\Delta$.

It should be noted that Eq.\ (\ref{eqn:awrtB}) characterizes the scattering
length near resonance only for purely elastic scattering. If there exist
lower-energy channels that allow decay, then $a(B)$ has a non-zero imaginary
component and does not follow the simple pole formula (\ref{eqn:awrtB})
\cite{Hutson:res:2007}. For $i_b=0$, the Hamiltonian (\ref{eqn:Ham}) allows
only elastic scattering even when the alkali-metal atom is in a magnetically
excited state. However, when $i_b \ne 0$, couplings involving $\hat
i_b\cdot\hat s$ can change $m_{f,a}$, and for alkali-metal atoms in
magnetically excited states this provides additional couplings to lower-lying
thresholds. We have previously described the behavior of $a(B)$ for resonances
in such states for the LiYb systems \cite{Brue:LiYb:2012}.

%%%%%%%%%%%%%%%%%%%%%%%%%%%%%%%%%%%%%%%%%%%%%%%%%%%%%%%%%%%%%%%%%%%%%%%%%%%%%%%%%%%%%%%%%%
\subsection{Resonance widths from Golden Rule} \label{sec:FGR}

Coupled-channel calculations of resonance widths are straightforward but
provide relatively little insight into the factors that affect resonance
widths. We therefore develop here an alternative approach based on Fermi's
Golden Rule that allows us to understand the factors that determine the widths.

Fermi's Golden Rule gives an expression for the width of a Feshbach resonance
in terms of the off-diagonal matrix element of $\hat H'$ (Eq.\
(\ref{eqn:Hprime})) between the bound state $|\alpha_2,m_{f,a},n\rangle$ (with
vibrational quantum number $n$) and the continuum state
$|\alpha_1,m_{f,a},k\rangle$ (labeled by wavevector $k$, where $E_{\rm
col}=\hbar^2k^2/2\mu$). The Breit-Wigner width in the energy domain,
$\Gamma_E$, is
\begin{equation}
\Gamma_E(k)=2\pi\left|\middle<\alpha_2,m_{f,a},n\middle|
\hat H'\middle|\alpha_1,m_{f,a},k\middle>\right|^2,
\label{eqn:goldenrule}
\end{equation}
where the continuum function is normalized to a $\delta$-function of energy and
has asymptotic amplitude $(2\mu/\pi\hbar^2 k)^{1/2}$. At limitingly low
collision energy, $\Gamma_E(k)$ behaves as \cite{RevModPhys.82.1225},
\begin{equation}
\Gamma_E(k) \xrightarrow{k\to 0} 2ka_{\rm bg}\Gamma_0
\label{eqn:gamma0}
\end{equation}
where $a_{\rm bg}$ is the same background scattering length as in Eq.\
(\ref{eqn:awrtB}). $\Gamma_0$ is independent of energy and is related to the
magnetic resonance width $\Delta$ of Eq.\ (\ref{eqn:awrtB}) by
\begin{equation}
\Delta=\frac{\Gamma_0}{\delta\mu_{\rm res}},
\label{eqn:GammaToDelta}
\end{equation}
where $\delta\mu_{\rm res}$ is the difference between the magnetic moment of
the molecular bound state and that of the free atom pair, which is simply the
difference in slope of the crossing lines in Figure \ref{fig:BreitRabi1}.

The expression for the magnetic resonance width $\Delta$ factorizes into
spin-dependent and radial terms,
\begin{equation}
\Delta = \frac{\pi I_{m_{f,a}}(B)^2 I_{nk}^2}{k a_{\rm bg}\delta\mu_{\rm res}},
\label{eqn:FGRDelta1}
\end{equation}
where
\begin{equation}
I_{m_{f,a}}(B)=\left<\alpha_2,m_{f,a}\right|\hat {i}_a\cdot\hat
s\left|\alpha_1,m_{f,a}\right>
\end{equation}
and
\begin{equation}
I_{nk} = \int_0^\infty\psi_{n}(R)\Delta\zeta_a(R)\psi_{k}(R)\,{\rm d}R.
\label{eqn:Ink}
\end{equation}

\begin{figure}
\includegraphics[width=\linewidth]{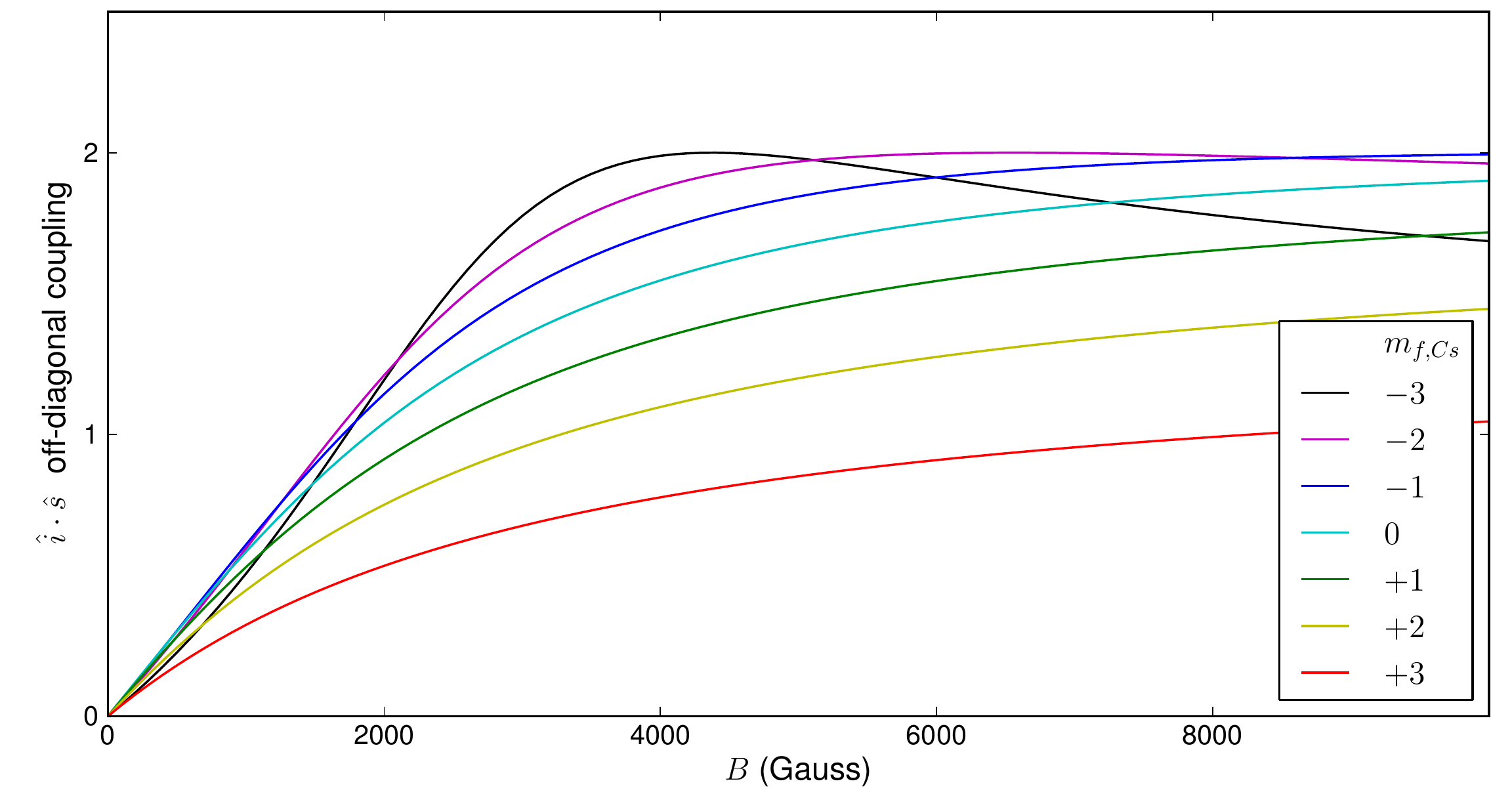}
\caption{(color online). Off-diagonal matrix elements of $\hat i_a\cdot\hat s$
for $^{133}$Cs between pairs of hyperfine states with the same value of
$m_{f,a}$.} \label{fig:idotsoffd}
\end{figure}

The quantity $I_{m_{f,a}}(B)$ is a purely atomic property, which arises because
states $|\alpha_i,m_{f,a}\rangle$ are eigenfunctions of $\hat H_a$. Pairs of
states with the same $m_{f,a}$ are coupled through the operator
$\Delta\zeta_a(R)\hat i_a \cdot\hat s$. At zero field, the states are
eigenfunctions of $\hat i_a\cdot\hat s$, so that the perturbation has no
off-diagonal matrix elements. At sufficiently high field, however, the states
are well described by quantum numbers $m_{s,a}$ and $m_{i,a}$, such that for a
given $m_{f,a}$ and $s_a=1/2$,
\begin{equation}
I_{m_{f,a}}(B) \xrightarrow{B\to\infty} \frac{1}{2}\left[i_a(i_a+1)-m_{f,a}^2+
\textstyle{\frac{1}{4}}\right]^\frac{1}{2}. \label{eq:islimit}
\end{equation}
The behavior of $I_{m_{f,a}}(B)$ between these two limits is shown as a
function of magnetic field for $^{133}$Cs in Figure \ref{fig:idotsoffd}. For
positive $m_{f,a}$ the coupling increases monotonically before leveling off to
the value (\ref{eq:islimit}), while for negative $m_{f,a}$ it increases with
$B$, peaks, and then declines to the same value. At low fields, the coupling is
approximately proportional to $B$, so that the resonance width is proportional
to $B^2$ in this region. The range over which this behavior occurs is
system-dependent; the coupling elements for lighter alkali metals level off at
smaller $B$ than for Cs.

The factor $1/\delta\mu_{\rm res}$ in Eq.\  (\ref{eqn:FGRDelta1}) produces
wider resonances when the difference in slope between the bound and continuum
states at $B_{\rm res}$ is small. Particularly shallow crossings and wide
resonances can occur when there is a ``double crossing" involving a bound state
that just dips below the threshold (as a function of $B$) before rising above
it again. The magnetic fields at which this can occur are discussed in Section
\ref{sec:results:choosing} below.

The bound and continuum functions, $\psi_n(R)$ and $\psi_k(R)$, are
eigenfunctions that correspond to different eigenvalues of the 1-dimensional
radial hamiltonian (\ref{eqn:Hrad}). They are thus orthogonal to one another,
and the matrix element $I_{nk}$ of Eq.\ (\ref{eqn:Ink}) is non-zero only
because of the $R$-dependence of $\Delta\zeta_a(R)$.

\begin{figure}
\begin{center}
\includegraphics[width=\linewidth]{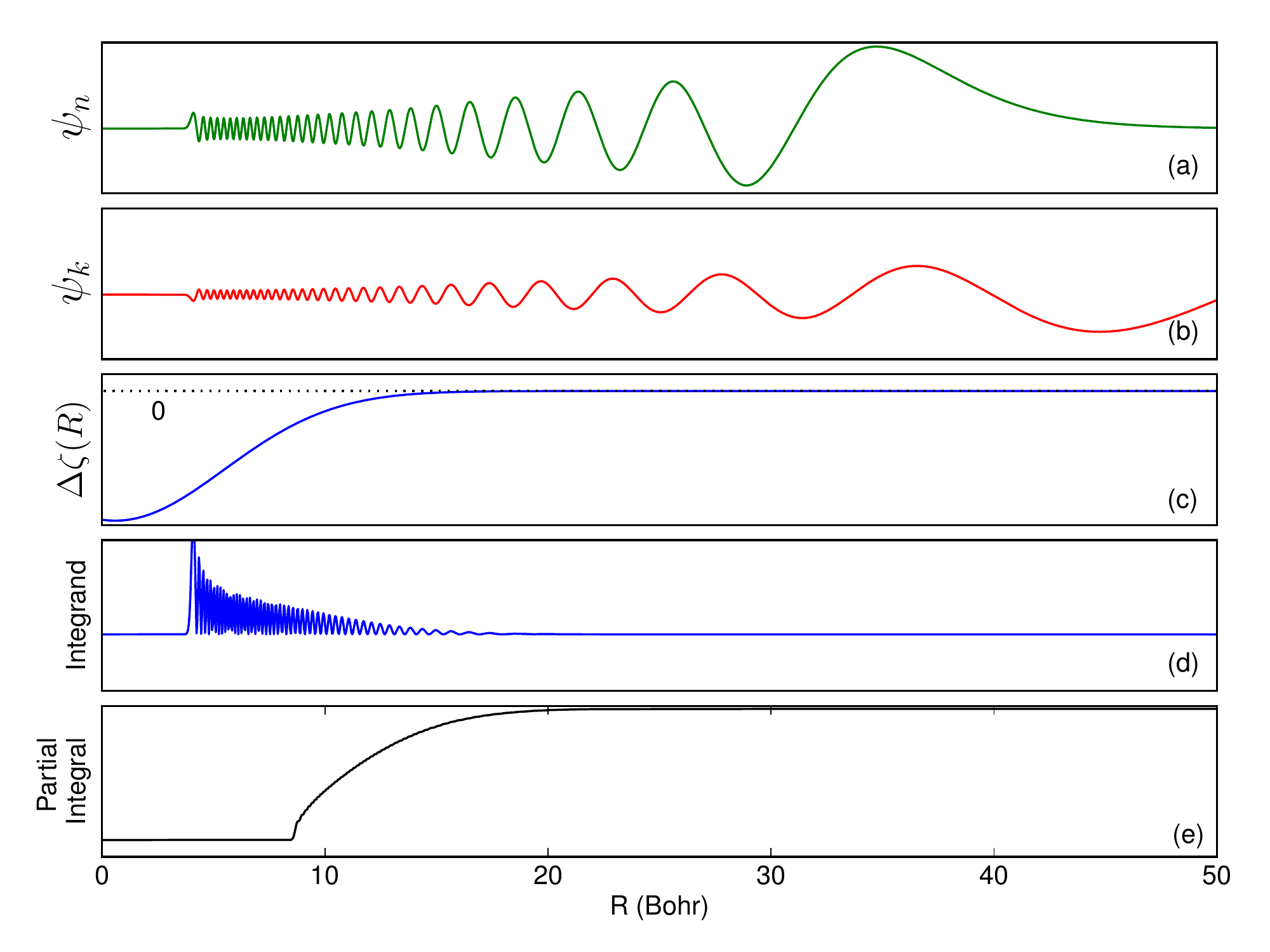}
\end{center}
\caption{(color online). The functions contributing to the integral $I_{nk}$
of Eq.\ (\ref{eqn:Ink}) for CsYb. (a) Wavefunction for the $n=-4$ bound state,
$\psi_{n=-4}(R)$, which is the shallowest bound state for which crossings exist.
(b) wavefunction for the low-energy continuum state, $\psi_k(R)$. (c)
$\Delta\zeta_{\rm Cs}(R)$. (d) the integrand of the matrix element,
$\psi_n\psi_k\Delta\zeta$. (e) The partial integral
$\int_0^R\psi_n(R^\prime)\psi_k(R^\prime)\Delta\zeta(R^\prime){\rm d}R^\prime$.
} \label{fig:CsYbIntegrand}
\end{figure}

Figure \ref{fig:CsYbIntegrand} shows how the integral $I_{nk}$ develops as a
function of $R$ in a typical case. The upper three panels show $\psi_n(R)$,
$\psi_k(R)$ and $\Delta\zeta_a(R)$. Figure \ref{fig:CsYbIntegrand}(d) shows the
integrand of Eq.\ (\ref{eqn:Ink}), which is the product of the three. For
weakly bound states, the bound and continuum functions remain almost in phase
with one another across the width of the potential well, so that their product
always maintains the same sign. The integral thus accumulates monotonically as
shown in the bottom panel. Its value depends principally on $\Delta\zeta_a(R)$
between the inner turning point and the potential minimum. Deeply bound states
lose phase with the continuum at shorter ranges; in principle this produces
some cancelation that reduces the value of the integral, but the effect of this
is small for the near-dissociation levels considered here.

Further insight may be gained by considering the integral $I_{nk}$
semiclassically. In the WKB (Wentzel-Kramers-Brillouin) approximation, the
bound and continuum wavefunctions both oscillate with amplitudes proportional
to $k(R)^{-1/2}$ in the classically allowed region, where $k(R) =
[2\mu(E-V(R))/\hbar^2]^{1/2}$. For very weakly bound states and low collision
energies, $E$ may be neglected, so
\begin{equation} I_{nk} \propto \int_{r_{\rm in}}^\infty
k(R)^{-1} \Delta\zeta(R)\,{\rm d}R,
\end{equation}
where $r_{\rm in}$ is the inner classical turning point at $E=0$. This
structure is clearly visible in Figure \ref{fig:CsYbIntegrand}(d).

Near threshold, the WKB approximation gives an incorrect ratio between the
short-range and long-range amplitudes of a scattering wavefunction. Quantum
Defect Theory (QDT) \cite{Mies:1984a} corrects for this using an
energy-dependent function $C(k)$, which is 1 far from threshold but is given by
\begin{equation}
C(k)^{-2} = k\bar{a}\left[1+\left(1-\frac{a_{\rm bg}}{\bar{a}}\right)^2\right]
\label{eqn:cqdt}
\end{equation}
at limitingly low energy \cite{RevModPhys.82.1225}. The correction amplifies
the short-range wavefunction by a factor $C(k)^{-1}$, which has a minimum value
of $(k\bar{a})^{1/2}$ when $a_{\rm bg}=\bar{a}$ but is approximately
$(k/\bar{a})^{1/2}a_{\rm bg}$ when $|a_{\rm bg}|\gg\bar{a}$.

Combining all these effects gives a semiclassical expression for the Golden
Rule width,
\begin{eqnarray}
\Delta &=& \frac{\mu}{\hbar^2} \frac{\bar{a}}{a_{\rm bg}}
\left[1+\left(1-\frac{a_{\rm bg}}{\bar{a}}\right)^2\right]
\frac{[I_{m_{f,a}}(B)]^2}{N\delta\mu_{\rm res}} \nonumber\\
&\times&
\left[\int_{r_{\rm in}}^\infty k(R)^{-1} \Delta\zeta(R)\,{\rm d}R\right]^2,
\label{eqn:FGRDeltaSC}
\end{eqnarray}
where $N$ is the normalization integral for the WKB bound-state wavefunction,
\begin{equation}
N = \frac{1}{2} \int_{R_{\rm in}}^{R_{\rm out}} k(R)^{-1}\,dR,
\label{eqn:kinv}
\end{equation}
which is taken between the classical turning points $R_{\rm in}$ and $R_{\rm
out}$ at energy $E_n$. Eq.\ (\ref{eqn:FGRDeltaSC}) completely avoids the
calculation of any quantal wavefunctions and gives results within 2\% of the
quantal Golden Rule width (\ref{eqn:FGRDelta1}).

The semiclassical approach may be taken one step further, with a small
approximation. For a near-dissociation vibrational state with an interaction
potential that varies as $-C_j R^{-j}$ at long range, Le Roy and Bernstein
\cite{LeRoy:1970} have shown that the integral (\ref{eqn:kinv}) is
\begin{equation}
\int_{R_{\rm in}}^{R_{\rm out}} k(R)^{-1}\,dR \approx
\left(\frac{\pi\hbar^2}{2\mu}\right)^{\frac{1}{2}}
\frac{\Gamma\left(\frac{1}{2}+\frac{1}{j}\right)}{\Gamma\left(1+\frac{1}{j}\right)}
\frac{C_j^{1/j}}{j} |E_n|^{-\frac{j+2}{2j}}, \label{eqn:kinvLB}
\end{equation}
where $\Gamma(x)$ is the gamma function. For the present case, with $R^{-6}$,
$\Delta$ is thus proportional to $|E_n|^{2/3}$. Deeper bound states thus
produce broader resonances, though generally at higher magnetic field. For the
bound states of interest here, Eq.\ (\ref{eqn:kinvLB}) is accurate to within
6\%.

As described below, different isotopes of Yb offer different values of the
scattering length $a_{\rm bg}$. Eq.\ (\ref{eqn:FGRDeltaSC}) shows that large
values of $\Delta$ may occur when $|a_{\rm bg}|$ is either very large or very
small: $\Delta$ is directly proportional to $a_{\rm bg}$ when $|a_{\rm
bg}|\gg\bar{a}$, and inversely proportional to $a_{\rm bg}$ when $|a_{\rm
bg}|\ll\bar{a}$.

Overall the Golden Rule approximation (\ref{eqn:FGRDeltaSC}) produces resonance
widths that agree within 2\% with those from full coupled-channel calculations.
It also produces important insights into the {\em origins} of the widths, and
makes it much easier to select systems and isotopic combinations with
experimentally desirable properties.

\subsection{Sensitivity to the interaction potential}
\label{sec:results:sensitivity}

The Feshbach resonance positions and widths are strongly dependent on the
s-wave scattering length of the system. The background scattering length
$a_{\rm bg}$, the binding energies of high-lying vibrational levels $E_n$, and
the non-integer quantum number at dissociation $v_{\rm D}$ can all be related
to a semiclassical phase integral $\Phi(E)$,
\begin{equation}
\Phi(E) = \int_{R_{\rm in}}^{R_{\rm out}} k(R)\,{\rm d}R.
\label{eqn:phaseint}
\end{equation}
For a potential with long-range behavior $V(R)=-C_6 R^{-6}$, the scattering
length is
\begin{equation}
a_{\rm bg} = \bar{a}\left[1-\tan\left(\Phi(0)-\frac{\pi}{8}\right)\right]
\label{eqn:abgbyabar}
\end{equation}
where $\bar{a}$ is the mean scattering length of Gribakin and Flambaum
\cite{GF}, which is proportional to $(\mu C_6)^{1/4}$. Values for $\bar{a}$ for
all the alkali metals with Yb atoms are given for representative isotopes in
Table \ref{tab:abar}. The non-integer quantum number at dissociation is
\begin{equation}
v_{\rm D}^{\rm GF} = \frac{\Phi(0)}{\pi}-\frac{5}{8},
\label{eqn:vdphi0}
\end{equation}
where the superscript GF distinguishes the Gribakin-Flambaum value from the
(less accurate) first-order WKB value (see section \ref{sec:results:choosing}).
It should be noted that $a_{\rm bg}$ is a single-valued function of the {\em
fractional part} of $v_{\rm D}^{\rm GF}$ and is independent of its integer
part.

\begin{table}
\begin{center}
\begin{tabular}{lrrrr}
\hline\hline
 &        $^{168}$Yb & $^{176}$Yb \\
\hline
$^6$Li    & 36.29 & 36.31 \\
$^7$Li    & 37.66 & 37.68 \\
$^{23}$Na &  50.50 &  50.57 \\
%$^{39}$K  & 62.77 & 62.91 \\
$^{40}$K  & 63.10 & 63.24 \\
%$^{41}$K  & 63.41 & 63.55 \\
%$^{85}$Rb & 74.24 & 74.52 \\
$^{87}$Rb & 74.52 & 74.82 \\
$^{133}$Cs & 83.05 & 83.48 \\
\hline\hline
\end{tabular}
\end{center}
\caption{Mean scattering lengths $\bar{a}$ (in bohr) for the Alk-Yb
systems.}\label{tab:abar}
\end{table}

Potential energy curves from electronic structure calculations for heavy
molecules are typically accurate to at best a few percent. For curves that
support 35 to 70 bound states, such as those for the systems considered here,
this uncertainty is enough to span more than 1 in $v_{\rm D}$. It is thus not
possible to predict $a_{\rm bg}$ for these systems from electronic structure
calculations alone. An experimental measurement is essential to limit the
possible range of $a_{\rm bg}$.

If the uncertainty in $v_{\rm D}$ is much greater than 1 and we assume that the
possible values of $\Phi(0)$ (and hence $v_{\rm D}$) are uniformly distributed
over such a range of uncertainty in $V(r)$, we find from Eq.\
(\ref{eqn:abgbyabar}) that there is a 50\% probability that $a_{\rm bg}$ is in
the range $[0,2\bar{a}]$, and a 70.5\% probability that it is in the range
$[-\bar{a},3\bar{a}]$.

Different isotopologues of the same molecule have different reduced masses
$\mu$. Since $k(R)$ is proportional to $\mu^{1/2}$, changing between different
isotopes of Yb alters $\Phi(0)$, and hence $v_{\rm D}$ and $a_{\rm bg}$, in a
very well-defined way, which depends only weakly on the potential well depth.
For the case of LiYb, changing the heavy-atom isotope has very little effect on
the reduced mass and therefore on $a_{\rm bg}$. For the heavier alkalis, by
contrast, changing the Yb isotope allows the scattering length to be tuned over
a wide range. Table \ref{tab:vD} summarizes the number of bound states and the
amount by which it may be tuned for all the alkali-metal + Yb systems.

\begin{table}
\begin{center}
\begin{tabular}{lrrrr}
\hline\hline
 &        $v_{\rm D}$($^{172}$Yb) & $\Delta v_{\rm D}$(Yb) \\
\hline
$^6$Li     & 23  & 0.02 \\
$^7$Li     & 25  & 0.02 \\
$^{23}$Na  & 35  & 0.10 \\
$^{40}$K   & 45  & 0.20 \\
$^{87}$Rb  & 62  & 0.49 \\
$^{133}$Cs & 69  & 0.70 \\
\hline\hline
\end{tabular}
\end{center}
\caption{The integer part of $v_{\rm D}$ for Alk-Yb systems, based on the
potential curves from CCSD(T) calculations, together with the amount $\Delta
v_{\rm D}$ by which $v_{\rm D}$ may be tuned by varying the isotope of Yb. Note
that the number of bound states is $v_{\rm D}+1$.}\label{tab:vD}
\end{table}

\section{Results and Discussion}
\label{sec:results} We have previously calculated resonance positions and
widths for the LiYb systems \cite{Brue:LiYb:2012}, using estimates of $a_{\rm
bg}$ obtained from thermalization measurements for $^6$Li$^{174}$Yb
\cite{Takahashi:LiYb, Gupta:LiYb}. In the following subsections, we present
calculations of resonance positions and widths for two cases representative of
the heavier alkali metals: RbYb, where the scattering lengths are approximately
known, and CsYb, where the scattering lengths have yet to be measured.

\subsection{RbYb}
\label{sec:results:rbyb} Interactions of RbYb mixtures have been studied by
G\"orlitz and coworkers \cite{Gorlitz:Rb87Yb174, Baumer:thesis:2010,
Muenchow:2011, Muenchow:thesis:2012}. Baumer {\em et al.}\
\cite{Gorlitz:Rb87Yb174, Baumer:thesis:2010} measured thermalization rates and
density profiles for mixtures of $^{87}$Rb with a variety of Yb isotopes, and
interpreted the results in terms of background scattering lengths. In
particular, $^{87}$Rb$^{174}$Yb was found to have an extremely large scattering
length, which produced phase separation of the atomic clouds, while
$^{87}$Rb$^{170}$Yb was found to have an extremely small one. M\"unchow {\em et
al.} \cite{Muenchow:2011, Muenchow:thesis:2012} measured 2-photon
photoassociation spectra of high-lying vibrational states of the electronic
ground state: for $^{87}$Rb$^{176}$Yb, 6 states were observed with binding
energies between about 300 MHz and 60 GHz \cite{Muenchow:2011,
Muenchow:thesis:2012}, whereas for each of $^{170}$Yb, $^{172}$Yb and
$^{174}$Yb, two states were observed with binding energies between 100 and 1500
MHz. M\"unchow \cite{Muenchow:thesis:2012} fitted the binding energies to a
Lennard-Jones potential model and inferred from the mass scaling that the
potential supports about 66 bound states for $^{87}$Rb$^{174}$Yb and
$^{87}$Rb$^{176}$Yb, with one fewer state for lighter Yb isotopes. The presence
of a bound state very close to dissociation in $^{87}$Rb$^{174}$Yb produces its
large positive scattering length.

The Lennard-Jones potential reproduces the experimental spectra satisfactorily,
but the mass scaling determines only the number of bound states and there is no
reason to expect the potential to have the correct well depth, equilibrium
distance, or inner turning point. These features are however important in the
calculation of resonance widths. We have therefore refitted the binding
energies measured by M\"unchow \cite{Muenchow:thesis:2012}, together with the
scattering length for $^{87}$Rb$^{170}$Yb, to obtain a new potential curve
based on our CCSD(T) results described above. Our best fit was obtained by
multiplying the CCSD(T) potential by a scaling factor $\lambda_{\rm
scl}=1.09581$ and adjusting $C_6$ to 2874.7~$E_{\rm h}a_0^6$, producing a
potential that supports 66 bound states for $^{87}$Rb$^{176}$Yb. We also
introduced long-range $C_8$ and $C_{10}$ coefficients related to $C_6$ by a
ratio $\gamma=C_8/C_6=C_{10}/C_8$, with an optimum value $\gamma=267.5$
bohr$^2$ for the potential above. Since the resulting long-range potential is
valid to shorter distances than the pure $C_6R^{-6}$ potential used for the
other systems, the switching function \cite{Janssen:2009} was applied between
20 and 30 bohr in this case. It should be noted that adequate fits could also
be obtained with one or two additional (or fewer) bound states: increasing
$\lambda_{\rm scl}$ by 0.036 and $C_6$ by $37\ E_{\rm h}a_0^6$ produces a
potential with one extra bound state at the bottom of the well but the
high-lying states almost unchanged.

\begin{figure}
\includegraphics[width=\linewidth]{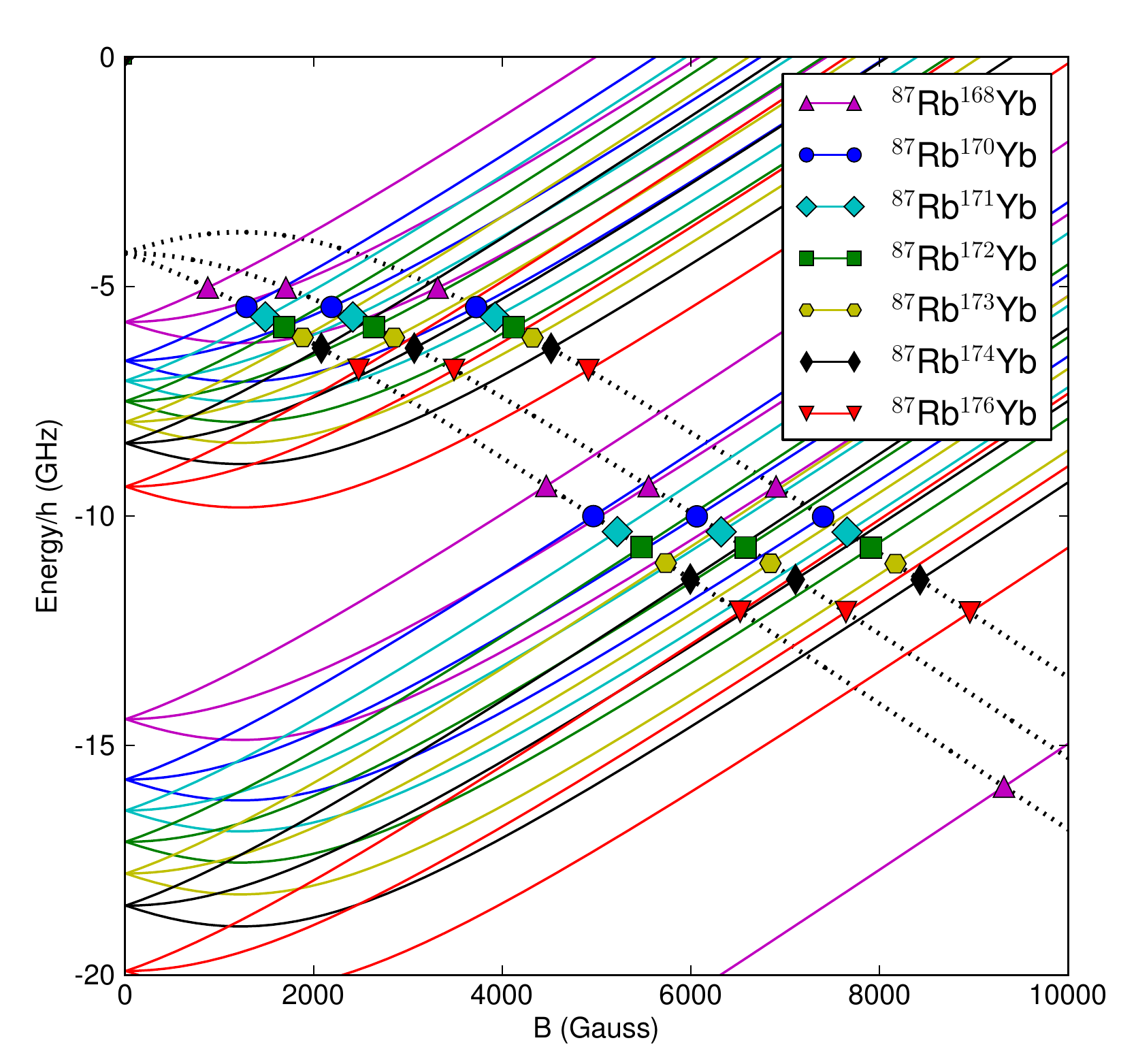}
\caption{(color online). Resonance crossings for $^{87}$Rb with the stable
isotopes of Yb, demonstrating the mass-scaling effect. The threshold levels for
$m_{f,{\rm Rb}}=-1,0,+1$ sublevels of the $f=1$ manifold are shown as dotted black lines,
while the molecular bound state $m_{f,{\rm Rb}}=-1,0,+1$ sublevels for the $f=2$ manifolds
are shown in different colors for the different isotopes of Yb. The bound
states for $m_{f,{\rm Rb}}=-2$ and +2 are not shown. The highest bound state shown here is
the $n=-4$ vibrational state for the combinations of $^{87}$Rb with
$^{168}$Yb$\to^{173}$Yb, and $n=-5$ for $^{87}$Rb$^{174}$Yb and
$^{87}$Rb$^{176}$Yb. } \label{fig:Rb87cross}
\end{figure}

\begin{figure}
\includegraphics[width=\linewidth]{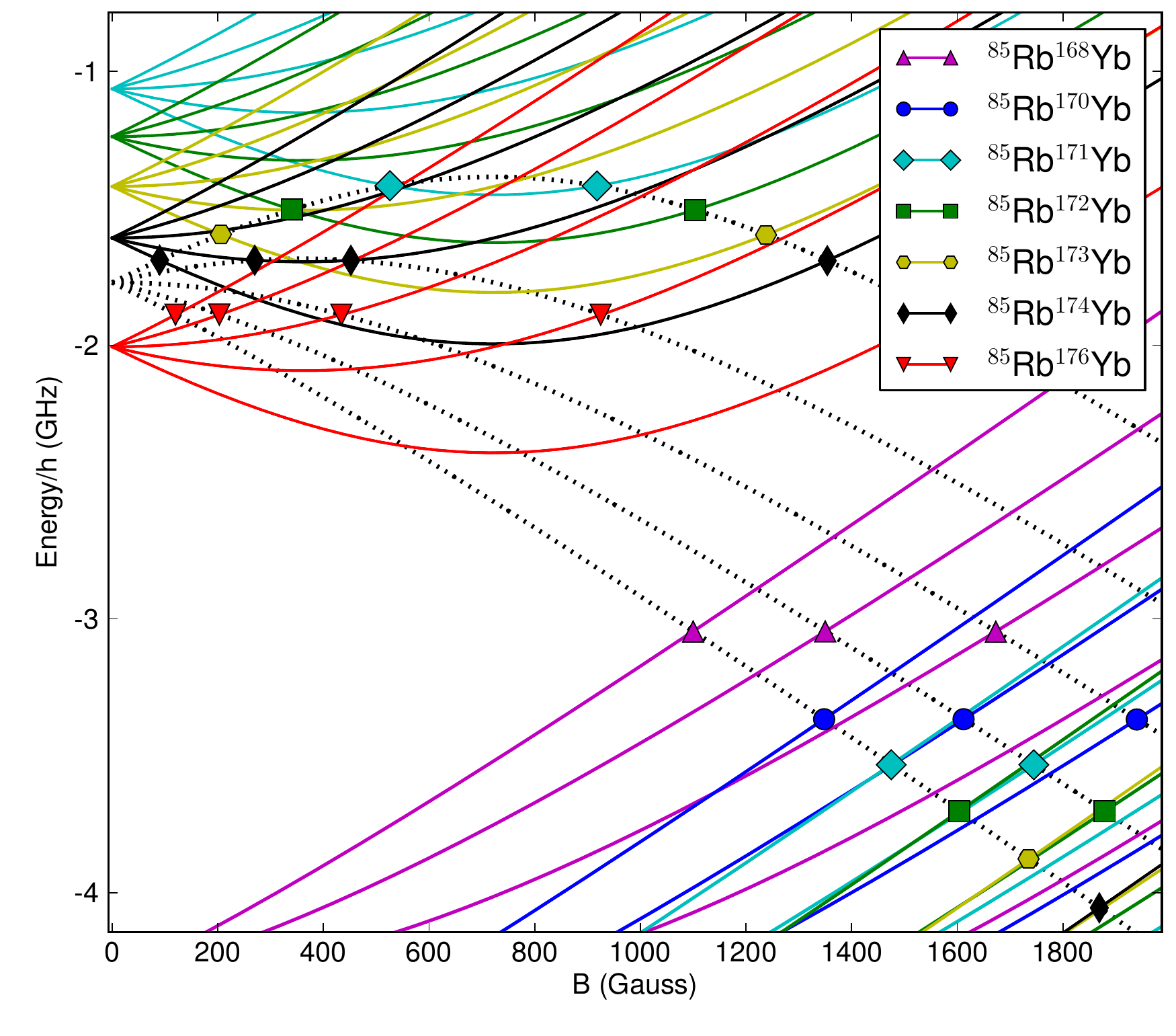}
\caption{(color online). Resonance crossings for $^{85}$Rb with the stable
isotopes of Yb. The threshold levels for $m_{f,{\rm Rb}}=-2,-1,0,+1,+2$ sublevels of the
$f=2$ manifold are shown as dotted black lines, while the corresponding
molecular bound-state sublevels for the $f=3$ manifolds are shown in different
colors for the different isotopes of Yb. The bound states for $m_{f,{\rm Rb}}=-3$ and +3
are not shown. The highest bound state shown here is the $n=-3$ vibrational
state.} \label{fig:Rb85cross}
\end{figure}

\begin{table}
\begin{center}
\begin{tabular}{lrrrrrr}
\hline\hline
Rb-Yb & $m_{f,{\rm Rb}}$  & $B_{\rm res}$ & $\Delta$ & $a_{\rm bg}$ & $\frac{|\Delta|}{B_{\rm res}}$ & $s_{\rm res}$\\
& & (G) & (mG) & (bohr) & (ppm) & \\
\hline
87-168 & $-1$ & 3314 &    4.6  &   39  &   1.4 & \quad $8.4\times10^{-4}$ \\
       &   0  & 1705 &    2.0  &   39  &   1.1 & $3.0\times10^{-4}$ \\
       &   1  &  877 &    0.3  &   39  &   0.4 & $5.9\times10^{-5}$ \\
       &   1  & 4466 &    2.4  &   39  &   0.5 & $5.9\times10^{-4}$ \\
\hline
87-170 & $-1$ & 3723 & $-31.1$ & $-11$ &   8.3 & $1.8\times10^{-3}$ \\
       &   0  & 2189 & $-16.6$ & $-11$ &   7.5 & $8.2\times10^{-4}$ \\
       &   1  & 1287 &  $-3.8$ & $-11$ &   2.9 & $2.1\times10^{-4}$ \\
       &   1  & 4965 & $-17.9$ & $-11$ &   3.6 & $1.3\times10^{-3}$ \\
\hline
87-171 & $-1$ & 3923 & $-10.9$ & $-58$ &   2.7 & $3.4\times10^{-3}$ \\
       &   0  & 2415 &  $-6.2$ & $-58$ &   2.5 & $1.7\times10^{-4}$ \\
       &   1  & 1487 &  $-1.6$ & $-58$ &   1.0 & $4.8\times10^{-4}$ \\
\hline
87-172 & $-1$ & 4122 & $-10.4$ &$-156$ &   2.5 & $8.9\times10^{-3}$ \\
       &   0  & 2636 &  $-6.2$ &$-156$ &   2.3 & $4.8\times10^{-3}$ \\
       &   1  & 1686 &  $-1.7$ &$-155$ &   1.0 & $1.5\times10^{-3}$ \\
\hline
87-173 & $-1$ & 4320 & $-20.7$ &$-576$ &   4.7 & $6.7\times10^{-2}$ \\
       &   0  & 2852 & $-13.0$ &$-571$ &   4.5 & $3.8\times10^{-2}$ \\
       &   1  & 1883 &  $-3.9$ &$-569$ &   2.0 & $1.3\times10^{-2}$ \\
\hline
87-174 & $-1$ & 4517 &   23.9  &  991  &   5.2 & $1.4\times10^{-1}$ \\
       &   0  & 3066 &   16.0  & 1000  &   5.2 & $8.5\times10^{-2}$ \\
       &   1  & 2081 &    5.2  & 1005  &   2.5 & $3.0\times10^{-2}$ \\
       \hline
87-176 & $-1$ & 4912 &    3.5  &  224  &   0.7 & $4.7\times10^{-3}$ \\
       &   0  & 3488 &    2.5  &  224  &   0.7 & $3.2\times10^{-3}$ \\
       &   1  & 2476 &    0.9  &  224  &   0.4 & $1.2\times10^{-3}$ \\
\hline\hline
\end{tabular}
\end{center}
\caption{Predicted positions and widths for resonances with $\Delta m_{f,{\rm
Rb}}=0$ for $^{87}$RbYb systems at fields $B_{\rm res} < 5000$~G.}
\label{tbl:Rb87Results}
\end{table}

\begin{table}
\begin{center}
\begin{tabular}{lrrrrrr}
\hline\hline
Rb-Yb & $m_{f,{\rm Rb}}$  & $B_{\rm res}$ & $\Delta$ & $a_{\rm bg}$ & $\frac{|\Delta|}{B_{\rm res}}$ & $s_{\rm res}$\\
& & (G) & (mG) & (bohr) & (ppm) & \\
\hline
85-168 & $ 1$ & 1350 &   0.17  &  219  &   0.12 & \quad $2.1\times10^{-4}$ \\
       & $ 2$ & 1100 &   0.066 &  219  &   0.06 &  $8.8\times10^{-5}$ \\
\hline
85-170 & $ 2$ & 1348 &   0.048 &  137  &   0.04 &  $4.1\times10^{-5}$ \\
\hline
85-171 & $-2$ &  526 & $-0.10$ &  116  &   0.18 &  $1.7\times10^{-5}$ \\
       & $-2$ &  918 &   0.30  &  116  &   0.32 &  $5.4\times10^{-4}$ \\
       & $ 2$ & 1475 &   0.048  &  116  &   0.03 & $4.1\times10^{-5}$ \\
\hline
85-172 & $-2$ &  340 & $-0.019$ &  99  &   0.06 &  $5.4\times10^{-6}$ \\
       & $-2$ & 1104 &   0.21  &   99  &   0.19 &  $6.1\times10^{-5}$ \\
\hline
85-173 & $-2$ &  206 & $-0.0055$&  84  &   0.03 &  $1.5\times10^{-6}$ \\
       & $-2$ & 1238 &   0.21  &   84  &   0.17 &  $6.4\times10^{-5}$ \\
\hline
85-174 & $-2$ &   90 & $-0.0009$&  70  &   0.01 &  $2.9\times10^{-7}$ \\
       & $-2$ & 1354 &   0.24  &   69  &   0.18 &  $7.0\times10^{-5}$ \\
       & $-1$ &  270 & $-0.11$ &   70  &   0.39 &  $4.4\times10^{-6}$ \\
       & $-1$ &  452 &   0.30  &   69  &   0.18 &  $1.2\times10^{-5}$ \\
\hline
85-176 & $-1$ &  925 &   0.43  &   39  &   0.46 &  $5.5\times10^{-5}$ \\
       &   0  &  434 &   0.14  &   39  &   0.32 &  $1.4\times10^{-5}$ \\
       &   1  &  203 &   0.021  &  39  &   0.10 &  $2.7\times10^{-6}$ \\
       &   2  &  120 &   0.0033 &  39  &   0.03 &  $5.7\times10^{-7}$ \\
\hline\hline
\end{tabular}
\end{center}
\caption{Predicted positions and widths for resonances with $\Delta m_{f,{\rm
Rb}}=0$ for $^{85}$RbYb systems at fields $B_{\rm res} < 1500$~G.}
\label{tbl:Rb85Results}
\end{table}

We have carried out coupled-channel calculations for the RbYb systems using the
fitted potential with 66 bound states. For the fermionic isotopes $^{171}$Yb
and $^{173}$Yb, we neglected couplings due to $\Delta\zeta_{\rm Yb}$. The
crossings responsible for the resonances for $^{87}$RbYb are shown in Figure
\ref{fig:Rb87cross} and the resonance positions and widths are given in Table
\ref{tbl:Rb87Results} for all resonances located below 5000 G. The
corresponding results for $^{85}$RbYb are given in Figure \ref{fig:Rb85cross}
and Table \ref{tbl:Rb85Results} for resonances located below 1500 G. A full
listing of all resonances below 10000~G is provided as Supplemental Material
\cite{SuppMatRbYb}. The resonance positions are generally within about 50~G of
those obtained by M\"unchow \cite{Muenchow:thesis:2012} with a Lennard-Jones
model of the potential.

The pattern of widths for $^{87}$RbYb closely follows expectations from Eq.\
(\ref{eqn:FGRDeltaSC}). Only $^{87}$Rb$^{168}$Yb has a resonance below 1000~G,
and that has a very low width (300~$\mu$G), in part because of dropoff in
$I_{m_{f,{\rm Rb}}}(B)$ at low fields. Nevertheless, resonances with calculated
widths as narrow as 0.2 $\mu$G have been observed as 3-body loss features in Na
\cite{Knoop:2011}, and resonances a few mG wide have been observed in LiNa at
fields as high as 2050~G \cite{Schuster:2012}. $^{87}$Rb$^{170}$Yb has
particularly large widths as measured by $\Delta$ (up to 30~mG), but this is
simply because $a_{\rm bg}$ is small in this case: the quantity $a_{\rm
bg}\Delta$, which is a better measure of the suitability of a resonance for
magnetoassociation \cite{Julienne:2004, Goral:2004}, is not particularly large
for this isotopologue. By contrast, $^{87}$Rb$^{173}$Yb and
$^{87}$Rb$^{174}$Yb, which both have $|a_{\rm bg}|\gg\bar{a}$, have resonances
up to 25~mG wide. Experimentally, $^{87}$Rb$^{174}$Yb displays phase separation
that will inhibit molecule formation even for low-temperature thermal clouds
\cite{Gorlitz:Rb87Yb174}, but $^{87}$Rb$^{173}$Yb does not
\cite{Baumer:thesis:2010}, and is a good candidate for magnetoassociation if
the high fields in Table \ref{tbl:Rb87Results} can be achieved.

For $^{85}$RbYb, there are no resonances with $\Delta/B_{\rm res} > 10^{-7}$.
This arises mostly because of the lower hyperfine coupling constant $\zeta$ for
$^{85}$Rb, which both reduces the magnitude of $\Delta\zeta(R)$ and further
reduces the widths through the factor of $|E_n|^{2/3}$ described following Eq.\
(\ref{eqn:kinvLB}) above. However, there are several resonances predicted below
1500~G, as shown in Table \ref{tbl:Rb85Results}, and some of the broader ones
(still below 1 mG width) may be suitable for molecule formation. In particular,
our best-fit potential predicts a pair of resonances for $m_{f,{\rm Rb}}=-1$
for $^{85}$Rb$^{174}$Yb, where the atomic and molecular states just intersect
and undergo a double crossing as shown in Fig.\ \ref{fig:Rb85cross}. The
precise positions and widths of these resonances are very sensitive to the
potential details, and indeed M\"unchow's Lennard-Jones model predicted that
the atomic and molecular states just miss each other instead of just crossing
\cite{Muenchow:thesis:2012}.

Tables \ref{tbl:Rb87Results} and \ref{tbl:Rb85Results} include only resonances
driven by $\Delta\zeta(R)$ for Rb, which conserve $m_{f,\rm Rb}$. If the Yb
isotope has nuclear spin, as for fermionic $^{171}$Yb and $^{173}$Yb,
additional resonances can occur at crossings with $\Delta m_{f,\rm Rb}=\pm1$,
driven by $\Delta\zeta(R)$ for Yb \cite{Brue:LiYb:2012}. In particular,
$^{87}$Rb$^{171}$Yb has a lower-field and therefore potentially more accessible
group of resonances near 1210~G, where the molecular states with $m_{f,{\rm
Rb}}=+2$ and $m_{i,{\rm Yb}}$ (not shown in Fig.\ \ref{fig:Rb87cross}) cross
the thresholds with $m_{f,{\rm Rb}}=+1$ and $m_{i,{\rm Yb}}-1$.

All the resonances in Tables \ref{tbl:Rb87Results} and \ref{tbl:Rb85Results}
are strongly closed-channel-dominated. This may be quantified using the
dimensionless resonance parameter $s_{\rm res} = (a_{\rm bg}/\bar{a})
(\delta\mu\Delta/\bar{E})$, where $\bar{E} = \hbar^2/(2\mu\bar{a}^2)$. It may
be seen that $s_{\rm res}$ is never greater than 0.2, and approaches such
values only when $|a_{\rm bg}|$ is very large. In some cases $s_{\rm res}$ can
be less than $10^{-6}$.

Molecule formation by magnetoassociation is usually carried out by preparing
the atomic mixture close to a resonance, on the side where the atomic state
lies below the molecular state, and then ramping the field over the resonance.
However, for narrow resonances in Cs$_2$ (a few mG wide, at low fields), Mark
{\em et al.}\ \cite{Mark:spect:2007} found it effective simply to hold the
field on resonance for a few milliseconds. Nevertheless, the most efficient
molecule production occurs with a field ramp that is slow enough to cross the
resonance adiabatically \cite{Julienne:2004, Goral:2004, Hodby:2005}. Small
field inhomogeneities are not a big problem, as they will simply cause
different parts of the cloud to cross the resonance at slightly different
times. However, field noise is potentially a problem, particularly
high-frequency noise that causes nonadiabatic crossings through the resonance.
It will therefore be important to design a molecule creation experiment with
very careful field control. In this context it is worth noting that Z\"urn {\em
et al.} \cite{Zurn:Li2-binding:2013} have recently carried out radiofrequency
spectroscopy on Li$_2$ molecules at fields around 800~G with a field precision
of $\pm1$~mG, which is close to 1 part in $10^6$, while Heo {\em et al.}
\cite{Heo:2012} achieved molecule formation in $^6$LiNa, using a resonance
10~mG wide at 745~G, with active feedback stabilization of the current to
achieve field noise less than 10~mG \cite{Heo:2012}.

\subsection{CsYb}
Cesium possesses several properties that make it favorable compared to the
other alkali-metal elements for magnetoassociation with Yb. It has the highest
mass of the alkali metals, which leads to greater mass scaling through changing
the isotope of the closed-shell atom. Its larger mass also provides a higher
density of bound states near threshold and thus offers better chances of
resonances at low magnetic field. Additionally, its relatively large nuclear
spin allows larger off-diagonal $\hat i_a \cdot \hat s$ elements. Finally, the
effects of $\Delta\zeta(R)$ are larger for Cs than for most of the other alkali
metals.

The CsYb potential shown in Fig.\ \ref{fig:AlkYbPots} supports 70 bound states
for all Yb isotopes, and has a background scattering length $a_{\rm bg}=-38$
bohr for $^{133}$Cs$^{174}$Yb. However, the electronic structure calculations
have a degree of inaccuracy, and a plausible change of $\pm10$\% in the well
depth would produce a change of $\pm3$ in $v_{\rm D}$. Since the scattering
length depends on the fractional part of $v_{\rm D}$, it cannot be predicted
from these calculations. However, altering the Yb isotopic mass across its
possible range from 168 to 176 changes $v_{\rm D}$ by about 0.70, so that a
wide range of background scattering lengths will be accessible by varying the
Yb isotope. We have therefore carried out calculations for CsYb as a function
of $a_{\rm bg}$.

\begin{figure}
\includegraphics[width=\linewidth]{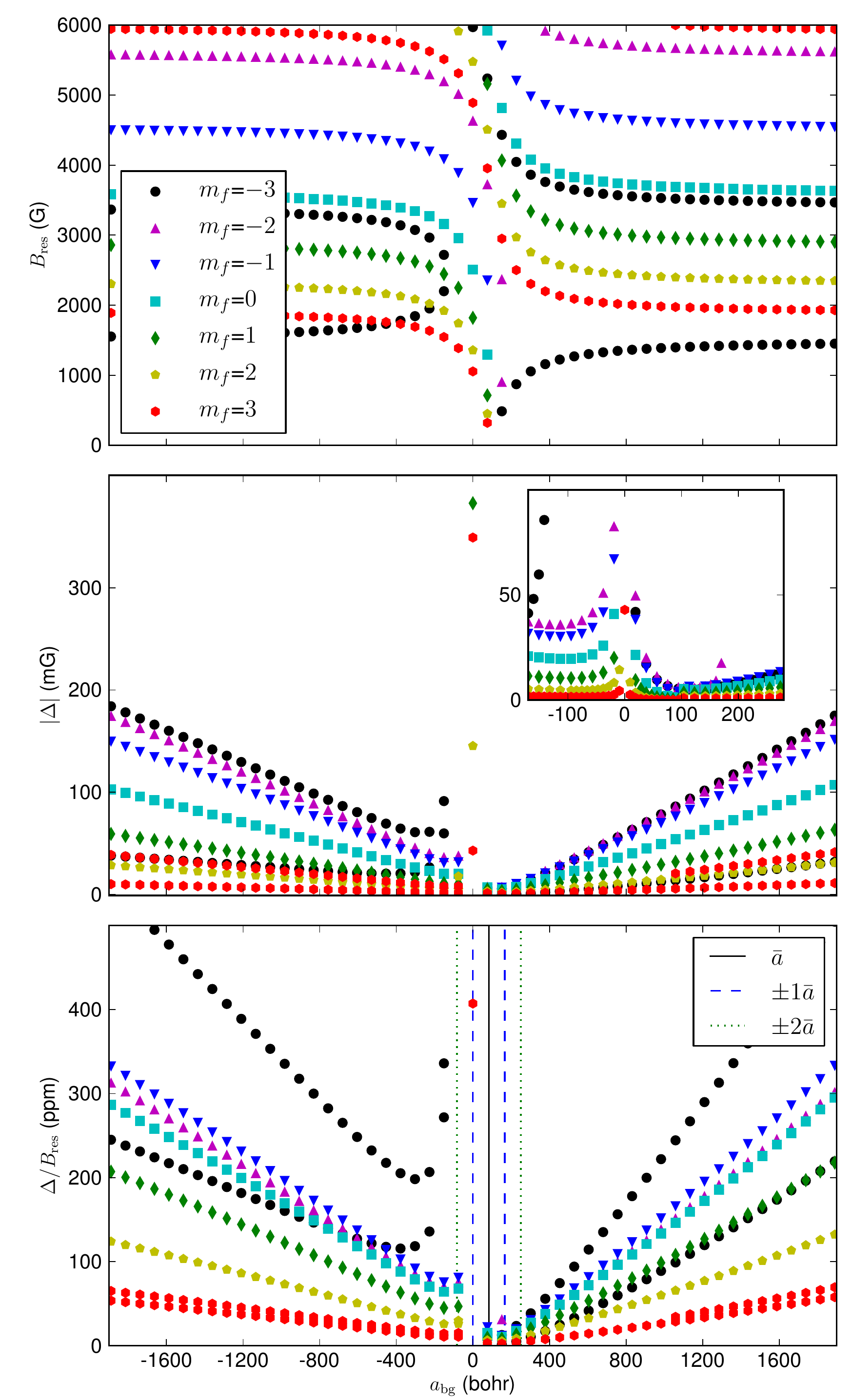}
\caption{(color online). Resonance positions and widths for
$^{133}$Cs$^{174}$Yb, obtained from coupled-channel calculations, as a function
of the background scattering length $a_{\rm bg}$. Top panel: Resonance position
$B_{\rm res}$. Center panel: Width $|\Delta|$, with an inset showing an
expanded view for values of $a_{\rm bg}$ near the mean scattering length,
$\bar{a}$. Bottom panel: $\Delta/B_{\rm res}$ in parts per million. }
\label{fig:CsYbAllRes}
\end{figure}

The top panel of Figure \ref{fig:CsYbAllRes} shows the resonance positions and
widths for $^{133}$Cs$^{174}$Yb as a function of $a_{\rm bg}$. The plot would
be almost identical for any other Yb isotope (though different isotopes will
have different scattering lengths). There are multiple resonances for each
value of $m_{f,{\rm Cs}}$, which occur when bound states $|\alpha_2,m_{f,{\rm
Cs}},n\rangle$ cross the scattering threshold $|\alpha_1,m_{f,{\rm
Cs}}\rangle$. As $a_{\rm bg}$ increases, the binding energies decrease and the
most of the crossings (those with positive $\delta\mu$) shift to lower magnetic
fields. In the $m_{f,{\rm Cs}}=+3$ case, the position of one resonance changes
from $B\approx 6000$ G to $B\approx 2000$ G as $a_{\rm bg}$ increases from
$-2000$ to $+2000$ bohr. When a state becomes too shallow to cross the lower
threshold at all, the corresponding resonance line either disappears through
$B=0$ or (in the case of a double crossing, as for $m_{f,{\rm Cs}}=-3$) reaches
a maximum $a_{\rm bg}$ where the two crossings coalesce.

The middle panel of Figure \ref{fig:CsYbAllRes} shows $|\Delta|$ as a function
of $a_{\rm bg}$. The spikes in $|\Delta|$ near $a_{\rm bg}=0$ occur because of
the $a_{\rm bg}$ in the denominator of Eq.\ (\ref{eqn:FGRDeltaSC}). However,
the strength of the peak in $a(B)$ [Eq.\ (\ref{eqn:awrtB})] is actually $a_{\rm
bg}\Delta$ rather than $\Delta$ itself, so this situation does not offer
particular advantages for molecule formation. For $|a_{\rm bg}\gg\bar{a}$, the
widths vary linearly with $a_{\rm bg}$ as described in Section \ref{sec:FGR}. A
particularly interesting feature of this plot is the spike in the $m_{f,{\rm
Cs}}=-3$ widths near $a_{\rm bg}=-100$ bohr, which is physically significant.
As noted above, the bound states for this magnetic sublevel experience a double
crossing with the lower threshold in this region; as the two crossings approach
one another, $\delta\mu_{\rm res}$ decreases and $\Delta$ increases as given by
Eq.\ (\ref{eqn:FGRDelta1}). A similar spike occurs in the $m_{f,{\rm Cs}}=-2$
resonance widths near 167 bohr. The inset shows an expanded view of $|\Delta|$
for the range of $a_{\rm bg}$ from $-\bar{a}$ to $3\bar{a}$; as described in
section \ref{sec:results:sensitivity}, there is about a 70\% probability that
$a_{\rm bg}$ lies in this range for any particular isotope.

%We have carried out similar coupled-channel calculations for CsHg. The results
%show resonance widths that are of the same order of magnitude as for CsYb. CsHg
%does have a larger reduced mass than CsYb which, all else being equal, would
%lead to a higher density of near-threshold bound states and thus more
%possibility of resonances at low  magnetic field. However, the potential for
%CsHg is only half as deep as that of CsYb, and so there is no marked increase
%in the number of near-threshold bound states. Furthermore, the larger mass
%disparity between Cs and Hg means that changing the Hg isotope does not shift
%the CsHg bound state energies as much as changing the Yb isotope does for CsYb.

\subsection{Choosing promising systems}
\label{sec:results:choosing} The Fermi Golden Rule treatment developed above
shows that the most important properties leading to large resonance widths are
a large magnitude of the background scattering length and the occurrence of
``double crossings" where the bound and continuum states have similar (small)
magnetic moments. It is instructive to consider the conditions where these two
enhancements can occur together.

At zero field, the hyperfine splitting of an alkali-metal atom in a $^2$S state
is $E_{\rm hf}(0)=\zeta\left(i+\frac{1}{2}\right)$. As a function of magnetic
field, the splitting between two states with the same value of $m_f$
(neglecting the nuclear Zeeman term) is
\begin{equation}
E_{\rm hf}(B) = \left[E_{\rm hf}(0)^2
+ \frac{4m_f E_{\rm hf}(0)}{2i+1} g_e \mu_{\rm B} B + (g_e\mu_{\rm B} B)^2 \right]^\frac{1}{2}.
\end{equation}
For negative values of $m_f$, this has a minimum value
\begin{equation}
E_{m_f}^{\rm close} = E_{\rm hf}(0)
\left( 1 - \frac{4m_f^2}{(2i+1)^2} \right)^\frac{1}{2}
\end{equation}
at a field
\begin{equation}
B_{m_f}^{\rm close} = \frac{-2m_f E_{\rm hf}(0)}{(2i+1) g_e \mu_{\rm B}}
=\frac{-m_f\zeta}{g_e \mu_{\rm B}}.
\end{equation}
The first-order WKB quantisation formula, expressed in terms of the phase
integral of Eq.\ (\ref{eqn:phaseint}), is
\begin{equation}
\Phi(E)=\left(v+\frac{1}{2}\right)\pi.
\label{eq:WKBquant}
\end{equation}
Le Roy and Bernstein \cite{LeRoy:1970} showed that this implies that, for a
long-range potential $V(R)=-C_j R^{-j}$ with $j>2$, near-dissociation levels
exist at energies
\begin{equation}
E_v = -[H_j (v_{\rm D}^{\rm WKB}-v)]^{2j/(j-2)},
\label{eq:LB}
\end{equation}
where
\begin{equation}
H_j = \left(\frac{\pi\hbar^2}{2\mu}\right)^\frac{1}{2} \frac{(j-2)}{C_j^{1/j}}
\frac{\Gamma\left(1+\frac{1}{j}\right)}{\Gamma\left(\frac{1}{2}+\frac{1}{j}\right)}.
\end{equation}
However, Eqs.\ (\ref{eq:WKBquant}) and (\ref{eq:LB}) do not take account of the
Gribakin-Flambaum correction \cite{GF}, which replaces the $(v+\frac{1}{2})$ in
Eq.\ (\ref{eq:WKBquant}) with $(v+\frac{1}{2}+\epsilon(v))$, where
$\epsilon(v)$ is zero for deeply bound levels but is $\frac{1}{8}$ at
dissociation for a long-range $R^{-6}$ potential. This correction may have a
significant effect on the energy of the least-bound level \cite{Boisseau:1998,
Boisseau:2000}, which is responsible for the Feshbach resonances Li-Yb
\cite{Brue:LiYb:2012} but is small for the slightly deeper levels that are
responsible for the resonances in the heavier Alk-Yb systems. As a result, the
near-dissociation levels (except $n=-1$) actually occur at energies close to
\begin{equation}
E_v = -H_6^3 \left(v_{\rm D}^{\rm WKB}-v\right)^3
= -H_6^3 \left(v_{\rm D}^{\rm GF}-v+\frac{1}{8}\right)^3,
\end{equation}
where \begin{equation} v_{\rm D}^{\rm
WKB}=\frac{\Phi(0)}{\pi}-\frac{1}{2}=v_{\rm D}^{\rm GF}+\frac{1}{8}
\end{equation}
when expressed in terms of $v_{\rm D}^{\rm GF}$ from Eq.\ (\ref{eqn:vdphi0}).

Very large values of $|a_{\rm bg}|$ correspond to near-integer values of
$v_{\rm D}^{\rm GF}$, so the condition for a very large value of $|a_{\rm bg}|$
to coexist with a ``double crossing" near $B_{m_f}^{\rm close}$ is that the
dimensionless quantity
\begin{equation}
X_{m_f} = \left[\frac{E_{\rm hf}(0)}{H_6^3}
\left( 1 - \frac{4m_f^2}{(2i+1)^2} \right)^\frac{1}{2}\right]^{1/3} - \frac{1}{8}
\label{eq:X}
\end{equation}
should be approximately an integer. The quantity $X_{m_f}$ may be interpreted
as the vibrational quantum number (relative to threshold) that will just give a
double crossing between a molecular state associated with the upper hyperfine
level and an atomic state at the lower hyperfine threshold with the same $m_f$.
It depends strongly on the alkali-metal isotope through the nuclear spin and
hyperfine splitting, but is only very weakly dependent on the Yb isotope
chosen. It is proportional to $C_6^{1/2}$ (through $H_6$), but is otherwise
completely independent of the interaction potential. Values of $X_{m_f}$
slightly smaller than than an integer allow a very large value of $|a_{\rm
bg}|$ to coexist with a double crossing further from $B_{m_f}^{\rm close}$, or
a large negative value of $a_{\rm bg}$ to coexist with a double crossing near
$B_{m_f}^{\rm close}$. In general, large negative values of $a_{\rm bg}$ may be
more favorable for molecule formation than large positive ones, because
negative values will not cause phase separation in condensates.

\begin{table}
\begin{center}
\begin{tabular}{lrrrrr}
\hline\hline
 & $m_f$ & $B_{m_f}^{\rm close}$ (G) & \quad $X_{m_f}$ & $a^{\max}_{m_f}$ (bohr) & $a^{\rm min}$ (bohr) \\
\hline
$^6$Li & $-1/2$    &  27 & 0.20 &  86 &  84 \\% && 0.14 &  83 &  81\\
$^7$Li & $-1$      & 143 & 0.40 &  51 &  47 \\% && 0.30 &  48 &  45\\
$^{23}$Na & $-1$   & 316 & 1.06 & 297 & 175 \\% && 0.86 &$-144$& $-94$\\
$^{39}$K  & $-1$   &  82 & 0.90 &$-122$ &$-302$ \\% && 0.71 & 11 & 0\\
$^{40}$K  & $-7/2$ & 357 & 1.17 & 173 &  87 \\% && 0.94 & $-194$ & 167 \\
$^{40}$K  & $-5/2$ & 255 & 1.29 & 111 &  87 \\% && 1.04 & 388 & 167\\
$^{40}$K  & $-3/2$ & 153 & 1.35 &  94 &  87 \\% && 1.09 & 201 & 167\\
$^{40}$K  & $-1/2$ &  51 & 1.38 &  88 &  87 \\% && 1.11 & 170 & 167\\
$^{41}$K  & $-1$   &  45 & 0.71 &  14 &  $-1$ \\% && 0.57 & 36 & 31\\
$^{85}$Rb & $-2$   & 722 & 2.36 &  111 & 48 \\% && 1.91 & $-135$ & 197\\
$^{85}$Rb & $-1$   & 361 & 2.56 &   61 & 48 \\% && 2.08 & 285 & 197\\
$^{87}$Rb & $-1$   &1219 & 3.32 &  123 & 78 \\% && 2.70 & 15 & $-45$ \\
$^{133}$Cs & $-3$  &2460 & 3.95 & $-399$ & 71 \\% && 3.21 & 139 & 14\\
$^{133}$Cs & $-2$ & 1640 & 4.33 &  133 & 71 \\% && 3.53 & 56 & 14\\
$^{133}$Cs & $-1$ &  820 & 4.50 &  84  & 71 \\% && 3.67 & 26 & 14\\
\hline\hline
\end{tabular}
\end{center}
\caption{The quantity $X_{m_f}$ of Eq.\ (\ref{eq:X}), which needs to be close
to an integer for double crossings to exist for large values of $|a_{\rm bg}|$,
together with the range of scattering lengths for which double crossings can
exist for a pure $R^{-6}$ potential. The values are almost independent of the
Yb isotope.}\label{tab:X}
\end{table}

Values of $X_{m_f}$ for all the Alk-Yb systems are given in Table \ref{tab:X}.
They may be converted into values of the scattering length that just cause
double crossings (for a pure $C_6/R^6$ potential) using
\begin{equation}
a^{\rm max}_{m_f} = \bar{a}
\left\{1-\tan\left[\pi\left(X_{m_f}+\textstyle{\frac{1}{2}}\right)\right]\right\}.
\label{eqn:amax}
\end{equation}
Scattering lengths between $a^{\rm max}_{m_f}$ and $a^{\rm min}=a^{\rm
max}_{0}$ will give rise to double crossings (where values of $a^{\rm
min}>a^{\rm max}_{m_f}$ are to be interpreted as allowing the scattering length
to be decreased from $a^{\rm max}_{m_f}$, through a pole and back down from
$+\infty$ to $a^{\rm min}$). However, only values close to $a^{\rm max}_{m_f}$
result in double crossings close to $B_{m_f}^{\rm close}$, which are the ones
with particularly large widths. Table \ref{tab:X} includes values of
$a_{m_f}^{\rm max}$ and $a^{\rm min}$ for all the Alk-Yb systems. It
immediately explains why $^{85}$Rb$^{174}$Yb, with a background scattering
length $a_{\rm bg}=70$ bohr that is reasonably close to $a^{\rm max}_{m_f}=61$
bohr, can have a double crossing near $B^{\rm close}=361$~G for $m_f=-1$. The
fact that this occurs with $a_{\rm bg}$ slightly larger than $a_{m_f}^{\rm
max}$ (rather than slightly smaller) reflects the approximations inherent in
Eq.\ (\ref{eqn:amax}): it applies only to a pure $C_6/R^6$ potential and only
approximately incorporates the Gribakin-Flambaum correction. Table \ref{tab:X}
also explains why Figure \ref{fig:CsYbAllRes} shows peaks in resonance widths
for CsYb at moderately large negative $a_{\rm bg}$ for $m_f=-3$ and for
moderately large positive $a_{\rm bg}$ for $m_f=-2$.

In general terms Yb is a favorable atom because it offers a large number of
isotopes that facilitate tuning the reduced mass and hence $a_{\rm bg}$. The
heavier alkali metals are more favorable than the light ones because their
larger masses offer greater tunability by varying the Yb mass. The heavier
alkali metals are also more favorable because the levels that offer crossings
at moderate magnetic fields have larger binding energies ($|E_n|$ between
$E_{\rm hf}^{\rm min}$ and $E_{\rm hf}(0)$). CsYb appears to be particularly
favorable because the near-integer value of $X_{-3}$ makes it possible for
shallow double crossings to coexist with large values of the scattering length.

As discussed above, the short-range amplitude of the bound-state wavefunction
is proportional to $|E_n|^{1/3}$. In addition, $\Delta\zeta_a(R)$ is very
roughly proportional to $\zeta_a$: for the Alk-Yb systems, $\zeta_0/\zeta_a$ is
about 0.3 for Li and Na and between 0.16 and 0.20 for K, Rb and Cs. The
integral $I_{nk}$ of Eq.\ (\ref{eqn:Ink}) thus scales very roughly as
$\zeta_a^{8/3}$ for resonances that occur at fields below $B^{\rm close}$. This
effect itself accounts for a factor of nearly 20 between the resonance widths
for $^{87}$Rb and $^{85}$Rb.
\section{Conclusion}
\label{sec:conclusion}

We have investigated Feshbach resonances in mixtures of alkali-metal atoms with
Yb, in order to identify promising systems for magnetoassociation to form
ultracold molecules with both electric and magnetic dipole moments. The
resonances in these systems arise when molecular states associated with the
upper hyperfine level of the alkali-metal atom cross atomic thresholds
associated with the lower hyperfine level. They are due to coupling by the
distance-dependence $\Delta\zeta(R)$ of the alkali-metal hyperfine coupling
constant \cite{PSZ:RbSr}. The widths of the resonances range from a few
microgauss to around 100 mG.

We have calculated the potential energy curves and $\Delta\zeta(R)$ for Yb
interacting with Na, K, Rb and Cs. We have carried out coupled-channel
calculations of the resonance positions and widths for all isotopologues of
RbYb and CsYb, and have also developed a perturbative model of the resonance
widths that gives good agreement with the coupled-channel results. Key
conclusions of the model are (i) that resonance widths depend strongly on the
atomic hyperfine coupling constant $\zeta$, with a general scaling as
$\zeta^{8/3}$; (ii) that resonance widths are generally proportional to the
background scattering length $a_{\rm bg}$ when it is larger than the mean
scattering length $\bar{a}$; (iii) that resonance widths are proportional to
$B^2$ in the low-field region where the atomic Zeeman effect is linear; (iv)
that unusually wide resonances may occur when a molecular bound state only just
crosses an atomic threshold as a function of $B$; (v) that, for the heavier
alkali metals, varying the Yb isotope gives access to a wide range of
background scattering lengths and thus to a range of different resonance
positions and properties. Selecting the best isotope is likely to be crucial to
the success of molecule production experiments.

Accurate predictions of resonance positions and widths for a given system
require knowledge of the background scattering length, or equivalently of the
binding energy of the least-bound vibrational state. This cannot be obtained
reliably from electronic structure calculations alone, and requires an
experimental measurement on at least one isotopologue. Once this is available,
the potential energy curves from electronic structure calculations are accurate
enough to allow mass-scaling to obtain predictions for {\em all} isotopologues.
For RbYb, for which binding energies have been measured by 2-photon
photoassociation spectroscopy \cite{Muenchow:thesis:2012}, we have adjusted our
potential curve to reproduce the experimental results and used the result to
calculate resonance positions and widths. We find that some isotopologues of
$^{85}$RbYb have resonances at fields below 1000~G, but these are all very
narrow ($<0.5$~mG). Isotopologues of $^{87}$RbYb have considerably wider
resonances (some up to 30~mG wide), but the most promising resonances occur at
fields above 2500~G.

For CsYb, no measurements of background scattering lengths or binding energies
are yet available. We have therefore calculated the resonance positions and
widths as a function of scattering length. CsYb is a particularly favorable
combination because shallow double crossings may occur for isotopologues with
large $a_{\rm bg}$, producing particularly broad resonances. The mapping from
scattering length to positions and widths is almost independent of
isotopologue, although the actual values of $a_{\rm bg}$ will be strongly
isotope-dependent.

\section*{Acknowledgments}
The authors are grateful to EPSRC for funding and to Piotr \.Zuchowski, Florian
Schreck, Axel G\"orlitz, Paul Julienne and Joe Cross for valuable discussions.

\bibliography{alk1s,../all}

\begin{thebibliography}{66}%
\makeatletter
\providecommand \@ifxundefined [1]{%
 \@ifx{#1\undefined}
}%
\providecommand \@ifnum [1]{%
 \ifnum #1\expandafter \@firstoftwo
 \else \expandafter \@secondoftwo
 \fi
}%
\providecommand \@ifx [1]{%
 \ifx #1\expandafter \@firstoftwo
 \else \expandafter \@secondoftwo
 \fi
}%
\providecommand \natexlab [1]{#1}%
\providecommand \enquote  [1]{``#1''}%
\providecommand \bibnamefont  [1]{#1}%
\providecommand \bibfnamefont [1]{#1}%
\providecommand \citenamefont [1]{#1}%
\providecommand \href@noop [0]{\@secondoftwo}%
\providecommand \href [0]{\begingroup \@sanitize@url \@href}%
\providecommand \@href[1]{\@@startlink{#1}\@@href}%
\providecommand \@@href[1]{\endgroup#1\@@endlink}%
\providecommand \@sanitize@url [0]{\catcode `\\12\catcode `\$12\catcode
  `\&12\catcode `\#12\catcode `\^12\catcode `\_12\catcode `\%12\relax}%
\providecommand \@@startlink[1]{}%
\providecommand \@@endlink[0]{}%
\providecommand \url  [0]{\begingroup\@sanitize@url \@url }%
\providecommand \@url [1]{\endgroup\@href {#1}{\urlprefix }}%
\providecommand \urlprefix  [0]{URL }%
\providecommand \Eprint [0]{\href }%
\providecommand \doibase [0]{http://dx.doi.org/}%
\providecommand \selectlanguage [0]{\@gobble}%
\providecommand \bibinfo  [0]{\@secondoftwo}%
\providecommand \bibfield  [0]{\@secondoftwo}%
\providecommand \translation [1]{[#1]}%
\providecommand \BibitemOpen [0]{}%
\providecommand \bibitemStop [0]{}%
\providecommand \bibitemNoStop [0]{.\EOS\space}%
\providecommand \EOS [0]{\spacefactor3000\relax}%
\providecommand \BibitemShut  [1]{\csname bibitem#1\endcsname}%
\let\auto@bib@innerbib\@empty
%</preamble>
\bibitem [{\citenamefont {Hudson}\ \emph {et~al.}(2006)\citenamefont {Hudson},
  \citenamefont {Lewandowski}, \citenamefont {Sawyer},\ and\ \citenamefont
  {Ye}}]{PhysRevLett.96.143004}%
  \BibitemOpen
  \bibfield  {author} {\bibinfo {author} {\bibfnamefont {E.~R.}\ \bibnamefont
  {Hudson}}, \bibinfo {author} {\bibfnamefont {H.~J.}\ \bibnamefont
  {Lewandowski}}, \bibinfo {author} {\bibfnamefont {B.~C.}\ \bibnamefont
  {Sawyer}}, \ and\ \bibinfo {author} {\bibfnamefont {J.}~\bibnamefont {Ye}},\
  }\href {\doibase 10.1103/PhysRevLett.96.143004} {\bibfield  {journal}
  {\bibinfo  {journal} {Phys. Rev. Lett.}\ }\textbf {\bibinfo {volume} {96}},\
  \bibinfo {pages} {143004} (\bibinfo {year} {2006})}\BibitemShut {NoStop}%
\bibitem [{\citenamefont {Kawall}\ \emph {et~al.}(2004)\citenamefont {Kawall},
  \citenamefont {Bay}, \citenamefont {Bickman}, \citenamefont {Jiang},\ and\
  \citenamefont {DeMille}}]{PhysRevLett.92.133007}%
  \BibitemOpen
  \bibfield  {author} {\bibinfo {author} {\bibfnamefont {D.}~\bibnamefont
  {Kawall}}, \bibinfo {author} {\bibfnamefont {F.}~\bibnamefont {Bay}},
  \bibinfo {author} {\bibfnamefont {S.}~\bibnamefont {Bickman}}, \bibinfo
  {author} {\bibfnamefont {Y.}~\bibnamefont {Jiang}}, \ and\ \bibinfo {author}
  {\bibfnamefont {D.}~\bibnamefont {DeMille}},\ }\href {\doibase
  10.1103/PhysRevLett.92.133007} {\bibfield  {journal} {\bibinfo  {journal}
  {Phys. Rev. Lett.}\ }\textbf {\bibinfo {volume} {92}},\ \bibinfo {pages}
  {133007} (\bibinfo {year} {2004})}\BibitemShut {NoStop}%
\bibitem [{\citenamefont {Hudson}\ \emph {et~al.}(2002)\citenamefont {Hudson},
  \citenamefont {Sauer}, \citenamefont {Tarbutt},\ and\ \citenamefont
  {Hinds}}]{PhysRevLett.89.023003}%
  \BibitemOpen
  \bibfield  {author} {\bibinfo {author} {\bibfnamefont {J.~J.}\ \bibnamefont
  {Hudson}}, \bibinfo {author} {\bibfnamefont {B.~E.}\ \bibnamefont {Sauer}},
  \bibinfo {author} {\bibfnamefont {M.~R.}\ \bibnamefont {Tarbutt}}, \ and\
  \bibinfo {author} {\bibfnamefont {E.~A.}\ \bibnamefont {Hinds}},\ }\href
  {\doibase 10.1103/PhysRevLett.89.023003} {\bibfield  {journal} {\bibinfo
  {journal} {Phys. Rev. Lett.}\ }\textbf {\bibinfo {volume} {89}},\ \bibinfo
  {pages} {023003} (\bibinfo {year} {2002})}\BibitemShut {NoStop}%
\bibitem [{\citenamefont {DeMille}(2002)}]{PhysRevLett.88.067901}%
  \BibitemOpen
  \bibfield  {author} {\bibinfo {author} {\bibfnamefont {D.}~\bibnamefont
  {DeMille}},\ }\href {\doibase 10.1103/PhysRevLett.88.067901} {\bibfield
  {journal} {\bibinfo  {journal} {Phys. Rev. Lett.}\ }\textbf {\bibinfo
  {volume} {88}},\ \bibinfo {pages} {067901} (\bibinfo {year}
  {2002})}\BibitemShut {NoStop}%
\bibitem [{\citenamefont {Jaksch}\ \emph {et~al.}(1999)\citenamefont {Jaksch},
  \citenamefont {Briegel}, \citenamefont {Cirac}, \citenamefont {Gardiner},\
  and\ \citenamefont {Zoller}}]{PhysRevLett.82.1975}%
  \BibitemOpen
  \bibfield  {author} {\bibinfo {author} {\bibfnamefont {D.}~\bibnamefont
  {Jaksch}}, \bibinfo {author} {\bibfnamefont {H.-J.}\ \bibnamefont {Briegel}},
  \bibinfo {author} {\bibfnamefont {J.~I.}\ \bibnamefont {Cirac}}, \bibinfo
  {author} {\bibfnamefont {C.~W.}\ \bibnamefont {Gardiner}}, \ and\ \bibinfo
  {author} {\bibfnamefont {P.}~\bibnamefont {Zoller}},\ }\href {\doibase
  10.1103/PhysRevLett.82.1975} {\bibfield  {journal} {\bibinfo  {journal}
  {Phys. Rev. Lett.}\ }\textbf {\bibinfo {volume} {82}},\ \bibinfo {pages}
  {1975} (\bibinfo {year} {1999})}\BibitemShut {NoStop}%
\bibitem [{\citenamefont {Micheli}\ \emph {et~al.}(2006)\citenamefont
  {Micheli}, \citenamefont {Brennen},\ and\ \citenamefont
  {Zoller}}]{Micheli:2006}%
  \BibitemOpen
  \bibfield  {author} {\bibinfo {author} {\bibfnamefont {A.}~\bibnamefont
  {Micheli}}, \bibinfo {author} {\bibfnamefont {G.~K.}\ \bibnamefont
  {Brennen}}, \ and\ \bibinfo {author} {\bibfnamefont {P.}~\bibnamefont
  {Zoller}},\ }\href@noop {} {\bibfield  {journal} {\bibinfo  {journal} {Nature
  Phys.}\ }\textbf {\bibinfo {volume} {2}},\ \bibinfo {pages} {341} (\bibinfo
  {year} {2006})}\BibitemShut {NoStop}%
\bibitem [{\citenamefont {Jones}\ \emph {et~al.}(2006)\citenamefont {Jones},
  \citenamefont {Tiesinga}, \citenamefont {Lett},\ and\ \citenamefont
  {Julienne}}]{RevModPhys.78.483}%
  \BibitemOpen
  \bibfield  {author} {\bibinfo {author} {\bibfnamefont {K.~M.}\ \bibnamefont
  {Jones}}, \bibinfo {author} {\bibfnamefont {E.}~\bibnamefont {Tiesinga}},
  \bibinfo {author} {\bibfnamefont {P.~D.}\ \bibnamefont {Lett}}, \ and\
  \bibinfo {author} {\bibfnamefont {P.~S.}\ \bibnamefont {Julienne}},\ }\href
  {\doibase 10.1103/RevModPhys.78.483} {\bibfield  {journal} {\bibinfo
  {journal} {Rev. Mod. Phys.}\ }\textbf {\bibinfo {volume} {78}},\ \bibinfo
  {pages} {483} (\bibinfo {year} {2006})}\BibitemShut {NoStop}%
\bibitem [{\citenamefont {K\"ohler}\ \emph {et~al.}(2006)\citenamefont
  {K\"ohler}, \citenamefont {G\'oral},\ and\ \citenamefont
  {Julienne}}]{RevModPhys.78.1311}%
  \BibitemOpen
  \bibfield  {author} {\bibinfo {author} {\bibfnamefont {T.}~\bibnamefont
  {K\"ohler}}, \bibinfo {author} {\bibfnamefont {K.}~\bibnamefont {G\'oral}}, \
  and\ \bibinfo {author} {\bibfnamefont {P.~S.}\ \bibnamefont {Julienne}},\
  }\href {\doibase 10.1103/RevModPhys.78.1311} {\bibfield  {journal} {\bibinfo
  {journal} {Rev. Mod. Phys.}\ }\textbf {\bibinfo {volume} {78}},\ \bibinfo
  {pages} {1311} (\bibinfo {year} {2006})}\BibitemShut {NoStop}%
\bibitem [{\citenamefont {Chin}\ \emph {et~al.}(2010)\citenamefont {Chin},
  \citenamefont {Grimm}, \citenamefont {Julienne},\ and\ \citenamefont
  {Tiesinga}}]{RevModPhys.82.1225}%
  \BibitemOpen
  \bibfield  {author} {\bibinfo {author} {\bibfnamefont {C.}~\bibnamefont
  {Chin}}, \bibinfo {author} {\bibfnamefont {R.}~\bibnamefont {Grimm}},
  \bibinfo {author} {\bibfnamefont {P.}~\bibnamefont {Julienne}}, \ and\
  \bibinfo {author} {\bibfnamefont {E.}~\bibnamefont {Tiesinga}},\ }\href
  {\doibase 10.1103/RevModPhys.82.1225} {\bibfield  {journal} {\bibinfo
  {journal} {Rev. Mod. Phys.}\ }\textbf {\bibinfo {volume} {82}},\ \bibinfo
  {pages} {1225} (\bibinfo {year} {2010})}\BibitemShut {NoStop}%
\bibitem [{\citenamefont {Herbig}\ \emph {et~al.}(2003)\citenamefont {Herbig},
  \citenamefont {Kraemer}, \citenamefont {Mark}, \citenamefont {Weber},
  \citenamefont {Chin}, \citenamefont {N\"agerl},\ and\ \citenamefont
  {Grimm}}]{Herbig:MolGas}%
  \BibitemOpen
  \bibfield  {author} {\bibinfo {author} {\bibfnamefont {J.}~\bibnamefont
  {Herbig}}, \bibinfo {author} {\bibfnamefont {T.}~\bibnamefont {Kraemer}},
  \bibinfo {author} {\bibfnamefont {M.}~\bibnamefont {Mark}}, \bibinfo {author}
  {\bibfnamefont {T.}~\bibnamefont {Weber}}, \bibinfo {author} {\bibfnamefont
  {C.}~\bibnamefont {Chin}}, \bibinfo {author} {\bibfnamefont {H.-C.}\
  \bibnamefont {N\"agerl}}, \ and\ \bibinfo {author} {\bibfnamefont
  {R.}~\bibnamefont {Grimm}},\ }\href@noop {} {\bibfield  {journal} {\bibinfo
  {journal} {Science}\ }\textbf {\bibinfo {volume} {301}},\ \bibinfo {pages}
  {1510} (\bibinfo {year} {2003})}\BibitemShut {NoStop}%
\bibitem [{\citenamefont {Regal}\ \emph {et~al.}(2003)\citenamefont {Regal},
  \citenamefont {Ticknor}, \citenamefont {Bohn},\ and\ \citenamefont
  {Jin}}]{Regal:MolGas}%
  \BibitemOpen
  \bibfield  {author} {\bibinfo {author} {\bibfnamefont {C.~A.}\ \bibnamefont
  {Regal}}, \bibinfo {author} {\bibfnamefont {C.}~\bibnamefont {Ticknor}},
  \bibinfo {author} {\bibfnamefont {J.~L.}\ \bibnamefont {Bohn}}, \ and\
  \bibinfo {author} {\bibfnamefont {D.~S.}\ \bibnamefont {Jin}},\ }\href@noop
  {} {\bibfield  {journal} {\bibinfo  {journal} {Nature}\ }\textbf {\bibinfo
  {volume} {424}},\ \bibinfo {pages} {47} (\bibinfo {year} {2003})}\BibitemShut
  {NoStop}%
\bibitem [{\citenamefont {Strecker}\ \emph {et~al.}(2003)\citenamefont
  {Strecker}, \citenamefont {Partridge},\ and\ \citenamefont
  {Hulet}}]{Strecker:MolGas}%
  \BibitemOpen
  \bibfield  {author} {\bibinfo {author} {\bibfnamefont {K.~E.}\ \bibnamefont
  {Strecker}}, \bibinfo {author} {\bibfnamefont {G.~B.}\ \bibnamefont
  {Partridge}}, \ and\ \bibinfo {author} {\bibfnamefont {R.~G.}\ \bibnamefont
  {Hulet}},\ }\href {\doibase 10.1103/PhysRevLett.91.080406} {\bibfield
  {journal} {\bibinfo  {journal} {Phys. Rev. Lett.}\ }\textbf {\bibinfo
  {volume} {91}},\ \bibinfo {pages} {080406} (\bibinfo {year}
  {2003})}\BibitemShut {NoStop}%
\bibitem [{\citenamefont {Cubizolles}\ \emph {et~al.}(2003)\citenamefont
  {Cubizolles}, \citenamefont {Bourdel}, \citenamefont {Kokkelmans},
  \citenamefont {Shlyapnikov},\ and\ \citenamefont
  {Salomon}}]{Cubizolles:MolGas}%
  \BibitemOpen
  \bibfield  {author} {\bibinfo {author} {\bibfnamefont {J.}~\bibnamefont
  {Cubizolles}}, \bibinfo {author} {\bibfnamefont {T.}~\bibnamefont {Bourdel}},
  \bibinfo {author} {\bibfnamefont {S.~J. J. M.~F.}\ \bibnamefont
  {Kokkelmans}}, \bibinfo {author} {\bibfnamefont {G.~V.}\ \bibnamefont
  {Shlyapnikov}}, \ and\ \bibinfo {author} {\bibfnamefont {C.}~\bibnamefont
  {Salomon}},\ }\href {\doibase 10.1103/PhysRevLett.91.240401} {\bibfield
  {journal} {\bibinfo  {journal} {Phys. Rev. Lett.}\ }\textbf {\bibinfo
  {volume} {91}},\ \bibinfo {pages} {240401} (\bibinfo {year}
  {2003})}\BibitemShut {NoStop}%
\bibitem [{\citenamefont {Jochim}\ \emph {et~al.}(2003)\citenamefont {Jochim},
  \citenamefont {Bartenstein}, \citenamefont {Altmeyer}, \citenamefont {Hendl},
  \citenamefont {Riedl}, \citenamefont {Chin}, \citenamefont
  {Hecker~Denschlag},\ and\ \citenamefont {Grimm}}]{Grimm:Li2BEC}%
  \BibitemOpen
  \bibfield  {author} {\bibinfo {author} {\bibfnamefont {S.}~\bibnamefont
  {Jochim}}, \bibinfo {author} {\bibfnamefont {M.}~\bibnamefont {Bartenstein}},
  \bibinfo {author} {\bibfnamefont {A.}~\bibnamefont {Altmeyer}}, \bibinfo
  {author} {\bibfnamefont {G.}~\bibnamefont {Hendl}}, \bibinfo {author}
  {\bibfnamefont {S.}~\bibnamefont {Riedl}}, \bibinfo {author} {\bibfnamefont
  {C.}~\bibnamefont {Chin}}, \bibinfo {author} {\bibfnamefont {J.}~\bibnamefont
  {Hecker~Denschlag}}, \ and\ \bibinfo {author} {\bibfnamefont
  {R.}~\bibnamefont {Grimm}},\ }\href@noop {} {\bibfield  {journal} {\bibinfo
  {journal} {Science}\ }\textbf {\bibinfo {volume} {302}},\ \bibinfo {pages}
  {2101} (\bibinfo {year} {2003})}\BibitemShut {NoStop}%
\bibitem [{\citenamefont {Ni}\ \emph {et~al.}(2008)\citenamefont {Ni},
  \citenamefont {Ospelkaus}, \citenamefont {de~Miranda}, \citenamefont {Pe'er},
  \citenamefont {Neyenhuis}, \citenamefont {Zirbel}, \citenamefont
  {Kotochigova}, \citenamefont {Julienne}, \citenamefont {Jin},\ and\
  \citenamefont {Ye}}]{Ni:KRb}%
  \BibitemOpen
  \bibfield  {author} {\bibinfo {author} {\bibfnamefont {K.-K.}\ \bibnamefont
  {Ni}}, \bibinfo {author} {\bibfnamefont {S.}~\bibnamefont {Ospelkaus}},
  \bibinfo {author} {\bibfnamefont {M.~H.~G.}\ \bibnamefont {de~Miranda}},
  \bibinfo {author} {\bibfnamefont {A.}~\bibnamefont {Pe'er}}, \bibinfo
  {author} {\bibfnamefont {B.}~\bibnamefont {Neyenhuis}}, \bibinfo {author}
  {\bibfnamefont {J.~J.}\ \bibnamefont {Zirbel}}, \bibinfo {author}
  {\bibfnamefont {S.}~\bibnamefont {Kotochigova}}, \bibinfo {author}
  {\bibfnamefont {P.~S.}\ \bibnamefont {Julienne}}, \bibinfo {author}
  {\bibfnamefont {D.~S.}\ \bibnamefont {Jin}}, \ and\ \bibinfo {author}
  {\bibfnamefont {J.}~\bibnamefont {Ye}},\ }\href@noop {} {\bibfield  {journal}
  {\bibinfo  {journal} {Science}\ }\textbf {\bibinfo {volume} {322}},\ \bibinfo
  {pages} {231} (\bibinfo {year} {10 Oct 2008})}\BibitemShut {NoStop}%
\bibitem [{\citenamefont {Danzl}\ \emph {et~al.}(2010)\citenamefont {Danzl},
  \citenamefont {Mark}, \citenamefont {Hallar}, \citenamefont {Gustavsson},
  \citenamefont {Hart}, \citenamefont {Aldegunde}, \citenamefont {Hutson},\
  and\ \citenamefont {N\"agerl}}]{Danzl:Cs2}%
  \BibitemOpen
  \bibfield  {author} {\bibinfo {author} {\bibfnamefont {J.~G.}\ \bibnamefont
  {Danzl}}, \bibinfo {author} {\bibfnamefont {M.~J.}\ \bibnamefont {Mark}},
  \bibinfo {author} {\bibfnamefont {E.}~\bibnamefont {Hallar}}, \bibinfo
  {author} {\bibfnamefont {M.}~\bibnamefont {Gustavsson}}, \bibinfo {author}
  {\bibfnamefont {R.}~\bibnamefont {Hart}}, \bibinfo {author} {\bibfnamefont
  {J.}~\bibnamefont {Aldegunde}}, \bibinfo {author} {\bibfnamefont {J.~M.}\
  \bibnamefont {Hutson}}, \ and\ \bibinfo {author} {\bibfnamefont {H.-C.}\
  \bibnamefont {N\"agerl}},\ }\href@noop {} {\bibfield  {journal} {\bibinfo
  {journal} {Nature Phys.}\ }\textbf {\bibinfo {volume} {6}},\ \bibinfo {pages}
  {265} (\bibinfo {year} {2010})}\BibitemShut {NoStop}%
\bibitem [{\citenamefont {Lang}\ \emph {et~al.}(2008)\citenamefont {Lang},
  \citenamefont {Winkler}, \citenamefont {Strauss}, \citenamefont {Grimm},\
  and\ \citenamefont {Hecker~Denschlag}}]{Lang:Rb2}%
  \BibitemOpen
  \bibfield  {author} {\bibinfo {author} {\bibfnamefont {F.}~\bibnamefont
  {Lang}}, \bibinfo {author} {\bibfnamefont {K.}~\bibnamefont {Winkler}},
  \bibinfo {author} {\bibfnamefont {C.}~\bibnamefont {Strauss}}, \bibinfo
  {author} {\bibfnamefont {R.}~\bibnamefont {Grimm}}, \ and\ \bibinfo {author}
  {\bibfnamefont {J.}~\bibnamefont {Hecker~Denschlag}},\ }\href {\doibase
  10.1103/PhysRevLett.101.133005} {\bibfield  {journal} {\bibinfo  {journal}
  {Phys. Rev. Lett.}\ }\textbf {\bibinfo {volume} {101}},\ \bibinfo {pages}
  {133005} (\bibinfo {year} {2008})}\BibitemShut {NoStop}%
\bibitem [{\citenamefont {Heo}\ \emph {et~al.}(2012)\citenamefont {Heo},
  \citenamefont {Wang}, \citenamefont {Christensen}, \citenamefont {Rvachov},
  \citenamefont {Cotta}, \citenamefont {Choi}, \citenamefont {Lee},\ and\
  \citenamefont {Ketterle}}]{Heo:2012}%
  \BibitemOpen
  \bibfield  {author} {\bibinfo {author} {\bibfnamefont {M.-S.}\ \bibnamefont
  {Heo}}, \bibinfo {author} {\bibfnamefont {T.~T.}\ \bibnamefont {Wang}},
  \bibinfo {author} {\bibfnamefont {C.~A.}\ \bibnamefont {Christensen}},
  \bibinfo {author} {\bibfnamefont {T.~M.}\ \bibnamefont {Rvachov}}, \bibinfo
  {author} {\bibfnamefont {D.~A.}\ \bibnamefont {Cotta}}, \bibinfo {author}
  {\bibfnamefont {J.-H.}\ \bibnamefont {Choi}}, \bibinfo {author}
  {\bibfnamefont {Y.-R.}\ \bibnamefont {Lee}}, \ and\ \bibinfo {author}
  {\bibfnamefont {W.}~\bibnamefont {Ketterle}},\ }\href@noop {} {\bibfield
  {journal} {\bibinfo  {journal} {Phys. Rev. A}\ }\textbf {\bibinfo {volume}
  {86}},\ \bibinfo {pages} {021602(R)} (\bibinfo {year} {2012})}\BibitemShut
  {NoStop}%
\bibitem [{\citenamefont {\.Zuchowski}\ \emph {et~al.}(2010)\citenamefont
  {\.Zuchowski}, \citenamefont {Aldegunde},\ and\ \citenamefont
  {Hutson}}]{PSZ:RbSr}%
  \BibitemOpen
  \bibfield  {author} {\bibinfo {author} {\bibfnamefont {P.~S.}\ \bibnamefont
  {\.Zuchowski}}, \bibinfo {author} {\bibfnamefont {J.}~\bibnamefont
  {Aldegunde}}, \ and\ \bibinfo {author} {\bibfnamefont {J.~M.}\ \bibnamefont
  {Hutson}},\ }\href {\doibase 10.1103/PhysRevLett.105.153201} {\bibfield
  {journal} {\bibinfo  {journal} {Phys. Rev. Lett.}\ }\textbf {\bibinfo
  {volume} {105}},\ \bibinfo {pages} {153201} (\bibinfo {year}
  {2010})}\BibitemShut {NoStop}%
\bibitem [{\citenamefont {Ivanov}\ \emph {et~al.}(2011)\citenamefont {Ivanov},
  \citenamefont {Khramov}, \citenamefont {Hansen}, \citenamefont {Dowd},
  \citenamefont {M\"unchow}, \citenamefont {Jamison},\ and\ \citenamefont
  {Gupta}}]{Gupta:LiYb}%
  \BibitemOpen
  \bibfield  {author} {\bibinfo {author} {\bibfnamefont {V.~V.}\ \bibnamefont
  {Ivanov}}, \bibinfo {author} {\bibfnamefont {A.}~\bibnamefont {Khramov}},
  \bibinfo {author} {\bibfnamefont {A.~H.}\ \bibnamefont {Hansen}}, \bibinfo
  {author} {\bibfnamefont {W.~H.}\ \bibnamefont {Dowd}}, \bibinfo {author}
  {\bibfnamefont {F.}~\bibnamefont {M\"unchow}}, \bibinfo {author}
  {\bibfnamefont {A.~O.}\ \bibnamefont {Jamison}}, \ and\ \bibinfo {author}
  {\bibfnamefont {S.}~\bibnamefont {Gupta}},\ }\href {\doibase
  10.1103/PhysRevLett.106.153201} {\bibfield  {journal} {\bibinfo  {journal}
  {Phys. Rev. Lett.}\ }\textbf {\bibinfo {volume} {106}},\ \bibinfo {pages}
  {153201} (\bibinfo {year} {2011})}\BibitemShut {NoStop}%
\bibitem [{\citenamefont {Hansen}\ \emph {et~al.}(2011)\citenamefont {Hansen},
  \citenamefont {Khramov}, \citenamefont {Dowd}, \citenamefont {Jamison},
  \citenamefont {Ivanov},\ and\ \citenamefont {Gupta}}]{Gupta:PRA:LiYb}%
  \BibitemOpen
  \bibfield  {author} {\bibinfo {author} {\bibfnamefont {A.~H.}\ \bibnamefont
  {Hansen}}, \bibinfo {author} {\bibfnamefont {A.}~\bibnamefont {Khramov}},
  \bibinfo {author} {\bibfnamefont {W.~H.}\ \bibnamefont {Dowd}}, \bibinfo
  {author} {\bibfnamefont {A.~O.}\ \bibnamefont {Jamison}}, \bibinfo {author}
  {\bibfnamefont {V.~V.}\ \bibnamefont {Ivanov}}, \ and\ \bibinfo {author}
  {\bibfnamefont {S.}~\bibnamefont {Gupta}},\ }\href {\doibase
  10.1103/PhysRevA.84.011606} {\bibfield  {journal} {\bibinfo  {journal} {Phys.
  Rev. A}\ }\textbf {\bibinfo {volume} {84}},\ \bibinfo {pages} {011606}
  (\bibinfo {year} {2011})}\BibitemShut {NoStop}%
\bibitem [{\citenamefont {Hara}\ \emph {et~al.}(2011)\citenamefont {Hara},
  \citenamefont {Takasu}, \citenamefont {Yamaoka}, \citenamefont {Doyle},\ and\
  \citenamefont {Takahashi}}]{Takahashi:LiYb}%
  \BibitemOpen
  \bibfield  {author} {\bibinfo {author} {\bibfnamefont {H.}~\bibnamefont
  {Hara}}, \bibinfo {author} {\bibfnamefont {Y.}~\bibnamefont {Takasu}},
  \bibinfo {author} {\bibfnamefont {Y.}~\bibnamefont {Yamaoka}}, \bibinfo
  {author} {\bibfnamefont {J.~M.}\ \bibnamefont {Doyle}}, \ and\ \bibinfo
  {author} {\bibfnamefont {Y.}~\bibnamefont {Takahashi}},\ }\href {\doibase
  10.1103/PhysRevLett.106.205304} {\bibfield  {journal} {\bibinfo  {journal}
  {Phys. Rev. Lett.}\ }\textbf {\bibinfo {volume} {106}},\ \bibinfo {pages}
  {205304} (\bibinfo {year} {2011})}\BibitemShut {NoStop}%
\bibitem [{\citenamefont {Nemitz}\ \emph {et~al.}(2009)\citenamefont {Nemitz},
  \citenamefont {Baumer}, \citenamefont {M\"unchow}, \citenamefont {Tassy},\
  and\ \citenamefont {G\"orlitz}}]{Gorlitz:RbYb}%
  \BibitemOpen
  \bibfield  {author} {\bibinfo {author} {\bibfnamefont {N.}~\bibnamefont
  {Nemitz}}, \bibinfo {author} {\bibfnamefont {F.}~\bibnamefont {Baumer}},
  \bibinfo {author} {\bibfnamefont {F.}~\bibnamefont {M\"unchow}}, \bibinfo
  {author} {\bibfnamefont {S.}~\bibnamefont {Tassy}}, \ and\ \bibinfo {author}
  {\bibfnamefont {A.}~\bibnamefont {G\"orlitz}},\ }\href {\doibase
  10.1103/PhysRevA.79.061403} {\bibfield  {journal} {\bibinfo  {journal} {Phys.
  Rev. A}\ }\textbf {\bibinfo {volume} {79}},\ \bibinfo {pages} {061403}
  (\bibinfo {year} {2009})}\BibitemShut {NoStop}%
\bibitem [{\citenamefont {Brue}\ and\ \citenamefont
  {Hutson}(2012)}]{Brue:LiYb:2012}%
  \BibitemOpen
  \bibfield  {author} {\bibinfo {author} {\bibfnamefont {D.~A.}\ \bibnamefont
  {Brue}}\ and\ \bibinfo {author} {\bibfnamefont {J.~M.}\ \bibnamefont
  {Hutson}},\ }\href@noop {} {\bibfield  {journal} {\bibinfo  {journal} {Phys.
  Rev. Lett.}\ }\textbf {\bibinfo {volume} {108}},\ \bibinfo {pages} {043201}
  (\bibinfo {year} {2012})}\BibitemShut {NoStop}%
\bibitem [{\citenamefont {Takasu}\ \emph {et~al.}(2003)\citenamefont {Takasu},
  \citenamefont {Maki}, \citenamefont {Komori}, \citenamefont {Takano},
  \citenamefont {Honda}, \citenamefont {Kumakura}, \citenamefont {Yabuzaki},\
  and\ \citenamefont {Takahashi}}]{Takasu:2003}%
  \BibitemOpen
  \bibfield  {author} {\bibinfo {author} {\bibfnamefont {Y.}~\bibnamefont
  {Takasu}}, \bibinfo {author} {\bibfnamefont {K.}~\bibnamefont {Maki}},
  \bibinfo {author} {\bibfnamefont {K.}~\bibnamefont {Komori}}, \bibinfo
  {author} {\bibfnamefont {T.}~\bibnamefont {Takano}}, \bibinfo {author}
  {\bibfnamefont {K.}~\bibnamefont {Honda}}, \bibinfo {author} {\bibfnamefont
  {M.}~\bibnamefont {Kumakura}}, \bibinfo {author} {\bibfnamefont
  {T.}~\bibnamefont {Yabuzaki}}, \ and\ \bibinfo {author} {\bibfnamefont
  {Y.}~\bibnamefont {Takahashi}},\ }\href@noop {} {\bibfield  {journal}
  {\bibinfo  {journal} {Phys. Rev. Lett.}\ }\textbf {\bibinfo {volume} {91}},\
  \bibinfo {pages} {040404} (\bibinfo {year} {2003})}\BibitemShut {NoStop}%
\bibitem [{\citenamefont {Fukuhara}\ \emph
  {et~al.}(2007{\natexlab{a}})\citenamefont {Fukuhara}, \citenamefont
  {Sugawa},\ and\ \citenamefont {Takahashi}}]{Fukuhara:boson:2007}%
  \BibitemOpen
  \bibfield  {author} {\bibinfo {author} {\bibfnamefont {T.}~\bibnamefont
  {Fukuhara}}, \bibinfo {author} {\bibfnamefont {S.}~\bibnamefont {Sugawa}}, \
  and\ \bibinfo {author} {\bibfnamefont {Y.}~\bibnamefont {Takahashi}},\
  }\href@noop {} {\bibfield  {journal} {\bibinfo  {journal} {Phys. Rev. A}\
  }\textbf {\bibinfo {volume} {76}},\ \bibinfo {pages} {051604} (\bibinfo
  {year} {2007}{\natexlab{a}})}\BibitemShut {NoStop}%
\bibitem [{\citenamefont {Fukuhara}\ \emph {et~al.}(2009)\citenamefont
  {Fukuhara}, \citenamefont {Sugawa}, \citenamefont {Takasu},\ and\
  \citenamefont {Takahashi}}]{Fukuhara:2009}%
  \BibitemOpen
  \bibfield  {author} {\bibinfo {author} {\bibfnamefont {T.}~\bibnamefont
  {Fukuhara}}, \bibinfo {author} {\bibfnamefont {S.}~\bibnamefont {Sugawa}},
  \bibinfo {author} {\bibfnamefont {Y.}~\bibnamefont {Takasu}}, \ and\ \bibinfo
  {author} {\bibfnamefont {Y.}~\bibnamefont {Takahashi}},\ }\href@noop {}
  {\bibfield  {journal} {\bibinfo  {journal} {Phys. Rev. A}\ }\textbf {\bibinfo
  {volume} {79}},\ \bibinfo {pages} {021601} (\bibinfo {year}
  {2009})}\BibitemShut {NoStop}%
\bibitem [{\citenamefont {Sugawa}\ \emph {et~al.}(2011)\citenamefont {Sugawa},
  \citenamefont {Yamazaki}, \citenamefont {Taie},\ and\ \citenamefont
  {Takahashi}}]{Sugawa:2011}%
  \BibitemOpen
  \bibfield  {author} {\bibinfo {author} {\bibfnamefont {S.}~\bibnamefont
  {Sugawa}}, \bibinfo {author} {\bibfnamefont {R.}~\bibnamefont {Yamazaki}},
  \bibinfo {author} {\bibfnamefont {S.}~\bibnamefont {Taie}}, \ and\ \bibinfo
  {author} {\bibfnamefont {Y.}~\bibnamefont {Takahashi}},\ }\href@noop {}
  {\bibfield  {journal} {\bibinfo  {journal} {Phys. Rev. A}\ }\textbf {\bibinfo
  {volume} {84}},\ \bibinfo {pages} {011610(R)} (\bibinfo {year}
  {2011})}\BibitemShut {NoStop}%
\bibitem [{\citenamefont {Fukuhara}\ \emph
  {et~al.}(2007{\natexlab{b}})\citenamefont {Fukuhara}, \citenamefont {Takasu},
  \citenamefont {Kumakura},\ and\ \citenamefont
  {Takahashi}}]{Fukuhara:Yb-173:2007}%
  \BibitemOpen
  \bibfield  {author} {\bibinfo {author} {\bibfnamefont {T.}~\bibnamefont
  {Fukuhara}}, \bibinfo {author} {\bibfnamefont {Y.}~\bibnamefont {Takasu}},
  \bibinfo {author} {\bibfnamefont {M.}~\bibnamefont {Kumakura}}, \ and\
  \bibinfo {author} {\bibfnamefont {Y.}~\bibnamefont {Takahashi}},\ }\href@noop
  {} {\bibfield  {journal} {\bibinfo  {journal} {Phys. Rev. Lett.}\ }\textbf
  {\bibinfo {volume} {98}},\ \bibinfo {pages} {030401} (\bibinfo {year}
  {2007}{\natexlab{b}})}\BibitemShut {NoStop}%
\bibitem [{\citenamefont {Fukuhara}\ \emph
  {et~al.}(2007{\natexlab{c}})\citenamefont {Fukuhara}, \citenamefont {Takasu},
  \citenamefont {Sugawa},\ and\ \citenamefont
  {Takahashi}}]{Fukuhara:fermion:2007}%
  \BibitemOpen
  \bibfield  {author} {\bibinfo {author} {\bibfnamefont {T.}~\bibnamefont
  {Fukuhara}}, \bibinfo {author} {\bibfnamefont {Y.}~\bibnamefont {Takasu}},
  \bibinfo {author} {\bibfnamefont {S.}~\bibnamefont {Sugawa}}, \ and\ \bibinfo
  {author} {\bibfnamefont {Y.}~\bibnamefont {Takahashi}},\ }\href@noop {}
  {\bibfield  {journal} {\bibinfo  {journal} {J. Low Temp. Phys.}\ }\textbf
  {\bibinfo {volume} {148}},\ \bibinfo {pages} {441} (\bibinfo {year}
  {2007}{\natexlab{c}})}\BibitemShut {NoStop}%
\bibitem [{\citenamefont {Strauss}\ \emph {et~al.}(2010)\citenamefont
  {Strauss}, \citenamefont {Takekoshi}, \citenamefont {Lang}, \citenamefont
  {Winkler}, \citenamefont {Grimm}, \citenamefont {Hecker~Denschlag},\ and\
  \citenamefont {Tiemann}}]{Strauss:2010}%
  \BibitemOpen
  \bibfield  {author} {\bibinfo {author} {\bibfnamefont {C.}~\bibnamefont
  {Strauss}}, \bibinfo {author} {\bibfnamefont {T.}~\bibnamefont {Takekoshi}},
  \bibinfo {author} {\bibfnamefont {F.}~\bibnamefont {Lang}}, \bibinfo {author}
  {\bibfnamefont {K.}~\bibnamefont {Winkler}}, \bibinfo {author} {\bibfnamefont
  {R.}~\bibnamefont {Grimm}}, \bibinfo {author} {\bibfnamefont
  {J.}~\bibnamefont {Hecker~Denschlag}}, \ and\ \bibinfo {author}
  {\bibfnamefont {E.}~\bibnamefont {Tiemann}},\ }\href@noop {} {\bibfield
  {journal} {\bibinfo  {journal} {Phys. Rev. A}\ }\textbf {\bibinfo {volume}
  {82}},\ \bibinfo {pages} {052514} (\bibinfo {year} {2010})}\BibitemShut
  {NoStop}%
\bibitem [{\citenamefont {Werner}\ \emph {et~al.}(2006)\citenamefont {Werner},
  \citenamefont {Knowles}, \citenamefont {Lindh}, \citenamefont {{Sch\"{u}tz}}
  \emph {et~al.}}]{MOLPRO_brief:2006}%
  \BibitemOpen
  \bibfield  {author} {\bibinfo {author} {\bibfnamefont {H.-J.}\ \bibnamefont
  {Werner}}, \bibinfo {author} {\bibfnamefont {P.~J.}\ \bibnamefont {Knowles}},
  \bibinfo {author} {\bibfnamefont {R.}~\bibnamefont {Lindh}}, \bibinfo
  {author} {\bibfnamefont {M.}~\bibnamefont {{Sch\"{u}tz}}},  \emph {et~al.},\
  }\href@noop {} {\enquote {\bibinfo {title} {{\sc MOLPRO}, version 2006.1: A
  package of ab initio programs},}\ } (\bibinfo {year} {2006}),\ \bibinfo
  {note} {see http://www.molpro.net}\BibitemShut {NoStop}%
\bibitem [{\citenamefont {Dolg}\ \emph {et~al.}(1989)\citenamefont {Dolg},
  \citenamefont {Stoll}, \citenamefont {Savin},\ and\ \citenamefont
  {Preuss}}]{DolgYbMWB}%
  \BibitemOpen
  \bibfield  {author} {\bibinfo {author} {\bibfnamefont {M.}~\bibnamefont
  {Dolg}}, \bibinfo {author} {\bibfnamefont {H.}~\bibnamefont {Stoll}},
  \bibinfo {author} {\bibfnamefont {A.}~\bibnamefont {Savin}}, \ and\ \bibinfo
  {author} {\bibfnamefont {H.}~\bibnamefont {Preuss}},\ }\href@noop {}
  {\bibfield  {journal} {\bibinfo  {journal} {Theoretical Chemistry Accounts}\
  }\textbf {\bibinfo {volume} {75}},\ \bibinfo {pages} {173} (\bibinfo {year}
  {1989})}\BibitemShut {NoStop}%
\bibitem [{\citenamefont {Leininger}\ \emph {et~al.}(1996)\citenamefont
  {Leininger}, \citenamefont {Nicklass}, \citenamefont {K\"achle},
  \citenamefont {Stoll}, \citenamefont {Dolg},\ and\ \citenamefont
  {Bergner}}]{Leininger1996274}%
  \BibitemOpen
  \bibfield  {author} {\bibinfo {author} {\bibfnamefont {T.}~\bibnamefont
  {Leininger}}, \bibinfo {author} {\bibfnamefont {A.}~\bibnamefont {Nicklass}},
  \bibinfo {author} {\bibfnamefont {W.}~\bibnamefont {K\"achle}}, \bibinfo
  {author} {\bibfnamefont {H.}~\bibnamefont {Stoll}}, \bibinfo {author}
  {\bibfnamefont {M.}~\bibnamefont {Dolg}}, \ and\ \bibinfo {author}
  {\bibfnamefont {A.}~\bibnamefont {Bergner}},\ }\href {\doibase
  10.1016/0009-2614(96)00382-X} {\bibfield  {journal} {\bibinfo  {journal}
  {Chemical Physics Letters}\ }\textbf {\bibinfo {volume} {255}},\ \bibinfo
  {pages} {274 } (\bibinfo {year} {1996})}\BibitemShut {NoStop}%
\bibitem [{\citenamefont {Stoll}()}]{StuttgartECPs}%
  \BibitemOpen
  \bibfield  {author} {\bibinfo {author} {\bibfnamefont {H.}~\bibnamefont
  {Stoll}},\ }\href@noop {} {\enquote {\bibinfo {title} {Pseudopotentials,
  {ECP}s},}\ }\bibinfo {howpublished}
  {http://www.theochem.uni-stuttgart.de/\~{}stoll/}\BibitemShut {NoStop}%
\bibitem [{\citenamefont {Prascher}\ \emph {et~al.}(2011)\citenamefont
  {Prascher}, \citenamefont {Woon}, \citenamefont {Peterson}, \citenamefont
  {{Dunning, Jr.}},\ and\ \citenamefont {Wilson}}]{NaBasis}%
  \BibitemOpen
  \bibfield  {author} {\bibinfo {author} {\bibfnamefont {B.}~\bibnamefont
  {Prascher}}, \bibinfo {author} {\bibfnamefont {D.~E.}\ \bibnamefont {Woon}},
  \bibinfo {author} {\bibfnamefont {K.~A.}\ \bibnamefont {Peterson}}, \bibinfo
  {author} {\bibfnamefont {T.~H.}\ \bibnamefont {{Dunning, Jr.}}}, \ and\
  \bibinfo {author} {\bibfnamefont {A.~K.}\ \bibnamefont {Wilson}},\
  }\href@noop {} {\bibfield  {journal} {\bibinfo  {journal} {Theoretical
  Chemistry Accounts}\ }\textbf {\bibinfo {volume} {128}},\ \bibinfo {pages}
  {69} (\bibinfo {year} {2011})}\BibitemShut {NoStop}%
\bibitem [{\citenamefont {Ho}\ and\ \citenamefont {Rabitz}(1995)}]{RKHS}%
  \BibitemOpen
  \bibfield  {author} {\bibinfo {author} {\bibfnamefont {T.-S.}\ \bibnamefont
  {Ho}}\ and\ \bibinfo {author} {\bibfnamefont {H.}~\bibnamefont {Rabitz}},\
  }\href@noop {} {\bibfield  {journal} {\bibinfo  {journal} {J. Chem. Phys.}\
  }\textbf {\bibinfo {volume} {104}},\ \bibinfo {pages} {2584} (\bibinfo {year}
  {1995})}\BibitemShut {NoStop}%
\bibitem [{\citenamefont {Zhang}\ \emph {et~al.}(2010)\citenamefont {Zhang},
  \citenamefont {Sadeghpour},\ and\ \citenamefont {Dalgarno}}]{Zhang:2010}%
  \BibitemOpen
  \bibfield  {author} {\bibinfo {author} {\bibfnamefont {P.}~\bibnamefont
  {Zhang}}, \bibinfo {author} {\bibfnamefont {H.~R.}\ \bibnamefont
  {Sadeghpour}}, \ and\ \bibinfo {author} {\bibfnamefont {A.}~\bibnamefont
  {Dalgarno}},\ }\href@noop {} {\bibfield  {journal} {\bibinfo  {journal} {J.
  Chem. Phys.}\ }\textbf {\bibinfo {volume} {133}},\ \bibinfo {pages} {044306}
  (\bibinfo {year} {2010})}\BibitemShut {NoStop}%
\bibitem [{\citenamefont {Tang}(1969)}]{PhysRev.177.108}%
  \BibitemOpen
  \bibfield  {author} {\bibinfo {author} {\bibfnamefont {K.~T.}\ \bibnamefont
  {Tang}},\ }\href {\doibase 10.1103/PhysRev.177.108} {\bibfield  {journal}
  {\bibinfo  {journal} {Phys. Rev.}\ }\textbf {\bibinfo {volume} {177}},\
  \bibinfo {pages} {108} (\bibinfo {year} {1969})}\BibitemShut {NoStop}%
\bibitem [{\citenamefont {Derevianko}\ \emph {et~al.}(2010)\citenamefont
  {Derevianko}, \citenamefont {Porsev},\ and\ \citenamefont
  {Babb}}]{Derevianko2010323}%
  \BibitemOpen
  \bibfield  {author} {\bibinfo {author} {\bibfnamefont {A.}~\bibnamefont
  {Derevianko}}, \bibinfo {author} {\bibfnamefont {S.~G.}\ \bibnamefont
  {Porsev}}, \ and\ \bibinfo {author} {\bibfnamefont {J.~F.}\ \bibnamefont
  {Babb}},\ }\href {\doibase DOI: 10.1016/j.adt.2009.12.002} {\bibfield
  {journal} {\bibinfo  {journal} {Atomic Data and Nuclear Data Tables}\
  }\textbf {\bibinfo {volume} {96}},\ \bibinfo {pages} {323 } (\bibinfo {year}
  {2010})}\BibitemShut {NoStop}%
\bibitem [{\citenamefont {Kitagawa}\ \emph {et~al.}(2008)\citenamefont
  {Kitagawa}, \citenamefont {Enomoto}, \citenamefont {Kasa}, \citenamefont
  {Takahashi}, \citenamefont {Ciury\l{}o}, \citenamefont {Naidon},\ and\
  \citenamefont {Julienne}}]{PhysRevA.77.012719}%
  \BibitemOpen
  \bibfield  {author} {\bibinfo {author} {\bibfnamefont {M.}~\bibnamefont
  {Kitagawa}}, \bibinfo {author} {\bibfnamefont {K.}~\bibnamefont {Enomoto}},
  \bibinfo {author} {\bibfnamefont {K.}~\bibnamefont {Kasa}}, \bibinfo {author}
  {\bibfnamefont {Y.}~\bibnamefont {Takahashi}}, \bibinfo {author}
  {\bibfnamefont {R.}~\bibnamefont {Ciury\l{}o}}, \bibinfo {author}
  {\bibfnamefont {P.}~\bibnamefont {Naidon}}, \ and\ \bibinfo {author}
  {\bibfnamefont {P.~S.}\ \bibnamefont {Julienne}},\ }\href {\doibase
  10.1103/PhysRevA.77.012719} {\bibfield  {journal} {\bibinfo  {journal} {Phys.
  Rev. A}\ }\textbf {\bibinfo {volume} {77}},\ \bibinfo {pages} {012719}
  (\bibinfo {year} {2008})}\BibitemShut {NoStop}%
\bibitem [{\citenamefont {Zhang}\ and\ \citenamefont
  {Dalgarno}(2007)}]{YbAlpha0}%
  \BibitemOpen
  \bibfield  {author} {\bibinfo {author} {\bibfnamefont {P.}~\bibnamefont
  {Zhang}}\ and\ \bibinfo {author} {\bibfnamefont {A.}~\bibnamefont
  {Dalgarno}},\ }\href@noop {} {\bibfield  {journal} {\bibinfo  {journal} {J.
  Phys. Chem A}\ }\textbf {\bibinfo {volume} {111}},\ \bibinfo {pages} {12471}
  (\bibinfo {year} {2007})}\BibitemShut {NoStop}%
\bibitem [{\citenamefont {Janssen}\ \emph {et~al.}(2009)\citenamefont
  {Janssen}, \citenamefont {Groenenboom}, \citenamefont {van~der Avoird},
  \citenamefont {\.Zuchowski},\ and\ \citenamefont {Podeszwa}}]{Janssen:2009}%
  \BibitemOpen
  \bibfield  {author} {\bibinfo {author} {\bibfnamefont {L.~M.~C.}\
  \bibnamefont {Janssen}}, \bibinfo {author} {\bibfnamefont {G.~C.}\
  \bibnamefont {Groenenboom}}, \bibinfo {author} {\bibfnamefont
  {A.}~\bibnamefont {van~der Avoird}}, \bibinfo {author} {\bibfnamefont
  {P.~S.}\ \bibnamefont {\.Zuchowski}}, \ and\ \bibinfo {author} {\bibfnamefont
  {R.}~\bibnamefont {Podeszwa}},\ }\href@noop {} {\bibfield  {journal}
  {\bibinfo  {journal} {J. Chem. Phys.}\ }\textbf {\bibinfo {volume} {131}},\
  \bibinfo {eid} {224314} (\bibinfo {year} {2009})}\BibitemShut {NoStop}%
\bibitem [{\citenamefont {Keal}\ and\ \citenamefont {Tozer}(2003)}]{Keal:2003}%
  \BibitemOpen
  \bibfield  {author} {\bibinfo {author} {\bibfnamefont {T.~W.}\ \bibnamefont
  {Keal}}\ and\ \bibinfo {author} {\bibfnamefont {D.~J.}\ \bibnamefont
  {Tozer}},\ }\href@noop {} {\bibfield  {journal} {\bibinfo  {journal} {J.
  Chem. Phys.}\ }\textbf {\bibinfo {volume} {119}},\ \bibinfo {pages} {3015}
  (\bibinfo {year} {2003})}\BibitemShut {NoStop}%
\bibitem [{ADF(2007)}]{ADF}%
  \BibitemOpen
  \href@noop {} {\enquote {\bibinfo {title} {{ADF2007.01}},}\ }\bibinfo
  {howpublished} {http://www.scm.com} (\bibinfo {year} {2007}),\ \bibinfo
  {note} {{SCM}, {T}heoretical {C}hemistry{,} {V}rije {U}niversiteit{,}
  {A}msterdam{,} {T}he {N}etherlands}\BibitemShut {NoStop}%
\bibitem [{\citenamefont {Moerdijk}\ \emph {et~al.}(1995)\citenamefont
  {Moerdijk}, \citenamefont {Verhaar},\ and\ \citenamefont
  {Axelsson}}]{PhysRevA.51.4852}%
  \BibitemOpen
  \bibfield  {author} {\bibinfo {author} {\bibfnamefont {A.~J.}\ \bibnamefont
  {Moerdijk}}, \bibinfo {author} {\bibfnamefont {B.~J.}\ \bibnamefont
  {Verhaar}}, \ and\ \bibinfo {author} {\bibfnamefont {A.}~\bibnamefont
  {Axelsson}},\ }\href {\doibase 10.1103/PhysRevA.51.4852} {\bibfield
  {journal} {\bibinfo  {journal} {Phys. Rev. A}\ }\textbf {\bibinfo {volume}
  {51}},\ \bibinfo {pages} {4852} (\bibinfo {year} {1995})}\BibitemShut
  {NoStop}%
\bibitem [{\citenamefont {Hutson}\ and\ \citenamefont
  {Green}(1994)}]{molscat:v14-short}%
  \BibitemOpen
  \bibfield  {author} {\bibinfo {author} {\bibfnamefont {J.~M.}\ \bibnamefont
  {Hutson}}\ and\ \bibinfo {author} {\bibfnamefont {S.}~\bibnamefont {Green}},\
  }\href@noop {} {\emph {\bibinfo {title} {{MOLSCAT} computer program, version
  14}}}\ (\bibinfo  {publisher} {CCP6},\ \bibinfo {address} {Daresbury},\
  \bibinfo {year} {1994})\BibitemShut {NoStop}%
\bibitem [{\citenamefont {Gonz\'{a}lez-Mart\'{\i}nez}\ and\ \citenamefont
  {Hutson}(2007)}]{Gonzalez-Martinez:2007}%
  \BibitemOpen
  \bibfield  {author} {\bibinfo {author} {\bibfnamefont {M.~L.}\ \bibnamefont
  {Gonz\'{a}lez-Mart\'{\i}nez}}\ and\ \bibinfo {author} {\bibfnamefont {J.~M.}\
  \bibnamefont {Hutson}},\ }\href@noop {} {\bibfield  {journal} {\bibinfo
  {journal} {Phys. Rev. A}\ }\textbf {\bibinfo {volume} {75}},\ \bibinfo
  {pages} {022702} (\bibinfo {year} {2007})}\BibitemShut {NoStop}%
\bibitem [{\citenamefont {Hutson}(2007)}]{Hutson:res:2007}%
  \BibitemOpen
  \bibfield  {author} {\bibinfo {author} {\bibfnamefont {J.~M.}\ \bibnamefont
  {Hutson}},\ }\href@noop {} {\bibfield  {journal} {\bibinfo  {journal} {New J.
  Phys.}\ }\textbf {\bibinfo {volume} {9}},\ \bibinfo {pages} {152} (\bibinfo
  {year} {2007})}\BibitemShut {NoStop}%
\bibitem [{\citenamefont {Mies}(1984)}]{Mies:1984a}%
  \BibitemOpen
  \bibfield  {author} {\bibinfo {author} {\bibfnamefont {F.~H.}\ \bibnamefont
  {Mies}},\ }\href {\doibase 10.1063/1.447000} {\bibfield  {journal} {\bibinfo
  {journal} {J. Chem. Phys.}\ }\textbf {\bibinfo {volume} {80}},\ \bibinfo
  {pages} {2514} (\bibinfo {year} {1984})}\BibitemShut {NoStop}%
\bibitem [{\citenamefont {Le~Roy}\ and\ \citenamefont
  {Bernstein}(1970)}]{LeRoy:1970}%
  \BibitemOpen
  \bibfield  {author} {\bibinfo {author} {\bibfnamefont {R.~J.}\ \bibnamefont
  {Le~Roy}}\ and\ \bibinfo {author} {\bibfnamefont {R.~B.}\ \bibnamefont
  {Bernstein}},\ }\href@noop {} {\bibfield  {journal} {\bibinfo  {journal} {J.
  Chem. Phys.}\ }\textbf {\bibinfo {volume} {52}},\ \bibinfo {pages} {3869}
  (\bibinfo {year} {1970})}\BibitemShut {NoStop}%
\bibitem [{\citenamefont {Gribakin}\ and\ \citenamefont {Flambaum}(1993)}]{GF}%
  \BibitemOpen
  \bibfield  {author} {\bibinfo {author} {\bibfnamefont {G.~F.}\ \bibnamefont
  {Gribakin}}\ and\ \bibinfo {author} {\bibfnamefont {V.~V.}\ \bibnamefont
  {Flambaum}},\ }\href {\doibase 10.1103/PhysRevA.48.546} {\bibfield  {journal}
  {\bibinfo  {journal} {Phys. Rev. A}\ }\textbf {\bibinfo {volume} {48}},\
  \bibinfo {pages} {546} (\bibinfo {year} {1993})}\BibitemShut {NoStop}%
\bibitem [{\citenamefont {Baumer}\ \emph {et~al.}(2011)\citenamefont {Baumer},
  \citenamefont {M\"unchow}, \citenamefont {G\"orlitz}, \citenamefont
  {Maxwell}, \citenamefont {Julienne},\ and\ \citenamefont
  {Tiesinga}}]{Gorlitz:Rb87Yb174}%
  \BibitemOpen
  \bibfield  {author} {\bibinfo {author} {\bibfnamefont {F.}~\bibnamefont
  {Baumer}}, \bibinfo {author} {\bibfnamefont {F.}~\bibnamefont {M\"unchow}},
  \bibinfo {author} {\bibfnamefont {A.}~\bibnamefont {G\"orlitz}}, \bibinfo
  {author} {\bibfnamefont {S.~E.}\ \bibnamefont {Maxwell}}, \bibinfo {author}
  {\bibfnamefont {P.~S.}\ \bibnamefont {Julienne}}, \ and\ \bibinfo {author}
  {\bibfnamefont {E.}~\bibnamefont {Tiesinga}},\ }\href {\doibase
  10.1103/PhysRevA.83.040702} {\bibfield  {journal} {\bibinfo  {journal} {Phys.
  Rev. A}\ }\textbf {\bibinfo {volume} {83}},\ \bibinfo {pages} {040702}
  (\bibinfo {year} {2011})}\BibitemShut {NoStop}%
\bibitem [{\citenamefont {Baumer}(2010)}]{Baumer:thesis:2010}%
  \BibitemOpen
  \bibfield  {author} {\bibinfo {author} {\bibfnamefont {F.}~\bibnamefont
  {Baumer}},\ }\emph {\bibinfo {title} {Isotope dependent interactions in a
  mixture of ultracold atoms}},\ \href@noop {} {Ph.D. thesis},\ \bibinfo
  {school} {Heinrich-Heine Universit\"at}, \bibinfo {address} {D\"usseldorf}
  (\bibinfo {year} {2010})\BibitemShut {NoStop}%
\bibitem [{\citenamefont {M\"unchow}\ \emph {et~al.}(2011)\citenamefont
  {M\"unchow}, \citenamefont {Bruni}, \citenamefont {Madalinskia},\ and\
  \citenamefont {G\"orlitz}}]{Muenchow:2011}%
  \BibitemOpen
  \bibfield  {author} {\bibinfo {author} {\bibfnamefont {F.}~\bibnamefont
  {M\"unchow}}, \bibinfo {author} {\bibfnamefont {C.}~\bibnamefont {Bruni}},
  \bibinfo {author} {\bibfnamefont {M.}~\bibnamefont {Madalinskia}}, \ and\
  \bibinfo {author} {\bibfnamefont {A.}~\bibnamefont {G\"orlitz}},\ }\href@noop
  {} {\bibfield  {journal} {\bibinfo  {journal} {Phys. Chem. Chem. Phys.}\
  }\textbf {\bibinfo {volume} {13}},\ \bibinfo {pages} {18734} (\bibinfo {year}
  {2011})}\BibitemShut {NoStop}%
\bibitem [{\citenamefont {M\"unchow}(2012)}]{Muenchow:thesis:2012}%
  \BibitemOpen
  \bibfield  {author} {\bibinfo {author} {\bibfnamefont {F.}~\bibnamefont
  {M\"unchow}},\ }\emph {\bibinfo {title} {2-photon photoassociation
  spectroscopy in a mixture of {Y}tterbium and {R}ubidium}},\ \href@noop {}
  {Ph.D. thesis},\ \bibinfo  {school} {Heinrich-Heine-Universit\"at}, \bibinfo
  {address} {D\"usseldorf} (\bibinfo {year} {2012})\BibitemShut {NoStop}%
\bibitem [{Sup()}]{SuppMatRbYb}%
  \BibitemOpen
  \href@noop {} {}\bibinfo {note} {See Supplemental Material at [URL will be
  inserted by publisher] for a full listing of all the predicted positions and
  widths of all resonances below 10000 G for $^{87}$RbYb and
  $^{85}$RbYb}\BibitemShut {NoStop}%
\bibitem [{\citenamefont {Knoop}\ \emph {et~al.}(2011)\citenamefont {Knoop},
  \citenamefont {Schuster}, \citenamefont {Scelle}, \citenamefont {Trautmann},
  \citenamefont {Appmeier}, \citenamefont {Oberthaler}, \citenamefont
  {Tiesinga},\ and\ \citenamefont {Tiemann}}]{Knoop:2011}%
  \BibitemOpen
  \bibfield  {author} {\bibinfo {author} {\bibfnamefont {S.}~\bibnamefont
  {Knoop}}, \bibinfo {author} {\bibfnamefont {T.}~\bibnamefont {Schuster}},
  \bibinfo {author} {\bibfnamefont {R.}~\bibnamefont {Scelle}}, \bibinfo
  {author} {\bibfnamefont {A.}~\bibnamefont {Trautmann}}, \bibinfo {author}
  {\bibfnamefont {J.}~\bibnamefont {Appmeier}}, \bibinfo {author}
  {\bibfnamefont {M.~K.}\ \bibnamefont {Oberthaler}}, \bibinfo {author}
  {\bibfnamefont {E.}~\bibnamefont {Tiesinga}}, \ and\ \bibinfo {author}
  {\bibfnamefont {E.}~\bibnamefont {Tiemann}},\ }\href@noop {} {\bibfield
  {journal} {\bibinfo  {journal} {Phys. Rev. A}\ }\textbf {\bibinfo {volume}
  {83}},\ \bibinfo {pages} {042704} (\bibinfo {year} {2011})}\BibitemShut
  {NoStop}%
\bibitem [{\citenamefont {Schuster}\ \emph {et~al.}(2012)\citenamefont
  {Schuster}, \citenamefont {Scelle}, \citenamefont {Trautmann}, \citenamefont
  {Knoop}, \citenamefont {Oberthaler}, \citenamefont {Haverhals}, \citenamefont
  {Goosen}, \citenamefont {Kokkelmans},\ and\ \citenamefont
  {Tiemann}}]{Schuster:2012}%
  \BibitemOpen
  \bibfield  {author} {\bibinfo {author} {\bibfnamefont {T.}~\bibnamefont
  {Schuster}}, \bibinfo {author} {\bibfnamefont {R.}~\bibnamefont {Scelle}},
  \bibinfo {author} {\bibfnamefont {A.}~\bibnamefont {Trautmann}}, \bibinfo
  {author} {\bibfnamefont {S.}~\bibnamefont {Knoop}}, \bibinfo {author}
  {\bibfnamefont {M.~K.}\ \bibnamefont {Oberthaler}}, \bibinfo {author}
  {\bibfnamefont {M.~M.}\ \bibnamefont {Haverhals}}, \bibinfo {author}
  {\bibfnamefont {M.~R.}\ \bibnamefont {Goosen}}, \bibinfo {author}
  {\bibfnamefont {S.~J. J. M.~F.}\ \bibnamefont {Kokkelmans}}, \ and\ \bibinfo
  {author} {\bibfnamefont {E.}~\bibnamefont {Tiemann}},\ }\href@noop {}
  {\bibfield  {journal} {\bibinfo  {journal} {Phys. Rev. A}\ }\textbf {\bibinfo
  {volume} {85}},\ \bibinfo {pages} {042721} (\bibinfo {year}
  {2012})}\BibitemShut {NoStop}%
\bibitem [{\citenamefont {Julienne}\ \emph {et~al.}(2004)\citenamefont
  {Julienne}, \citenamefont {{T}iesinga},\ and\ \citenamefont
  {K\"ohler}}]{Julienne:2004}%
  \BibitemOpen
  \bibfield  {author} {\bibinfo {author} {\bibfnamefont {P.~S.}\ \bibnamefont
  {Julienne}}, \bibinfo {author} {\bibfnamefont {E.}~\bibnamefont
  {{T}iesinga}}, \ and\ \bibinfo {author} {\bibfnamefont {T.}~\bibnamefont
  {K\"ohler}},\ }\href@noop {} {\bibfield  {journal} {\bibinfo  {journal} {J.
  Mod. Opt.}\ }\textbf {\bibinfo {volume} {51}},\ \bibinfo {pages} {1787}
  (\bibinfo {year} {2004})}\BibitemShut {NoStop}%
\bibitem [{\citenamefont {G\'{o}ral}\ \emph {et~al.}(2004)\citenamefont
  {G\'{o}ral}, \citenamefont {K\"ohler}, \citenamefont {Gardiner},
  \citenamefont {Tiesinga},\ and\ \citenamefont {Julienne}}]{Goral:2004}%
  \BibitemOpen
  \bibfield  {author} {\bibinfo {author} {\bibfnamefont {K.}~\bibnamefont
  {G\'{o}ral}}, \bibinfo {author} {\bibfnamefont {T.}~\bibnamefont {K\"ohler}},
  \bibinfo {author} {\bibfnamefont {S.~A.}\ \bibnamefont {Gardiner}}, \bibinfo
  {author} {\bibfnamefont {E.}~\bibnamefont {Tiesinga}}, \ and\ \bibinfo
  {author} {\bibfnamefont {P.~S.}\ \bibnamefont {Julienne}},\ }\href@noop {}
  {\bibfield  {journal} {\bibinfo  {journal} {J. Phys. B}\ }\textbf {\bibinfo
  {volume} {37}},\ \bibinfo {pages} {3427} (\bibinfo {year}
  {2004})}\BibitemShut {NoStop}%
\bibitem [{\citenamefont {Mark}\ \emph {et~al.}(2007)\citenamefont {Mark},
  \citenamefont {Ferlaino}, \citenamefont {Knoop}, \citenamefont {Danzl},
  \citenamefont {Kraemer}, \citenamefont {Chin}, \citenamefont {N\"agerl},\
  and\ \citenamefont {Grimm}}]{Mark:spect:2007}%
  \BibitemOpen
  \bibfield  {author} {\bibinfo {author} {\bibfnamefont {M.}~\bibnamefont
  {Mark}}, \bibinfo {author} {\bibfnamefont {F.}~\bibnamefont {Ferlaino}},
  \bibinfo {author} {\bibfnamefont {S.}~\bibnamefont {Knoop}}, \bibinfo
  {author} {\bibfnamefont {J.~G.}\ \bibnamefont {Danzl}}, \bibinfo {author}
  {\bibfnamefont {T.}~\bibnamefont {Kraemer}}, \bibinfo {author} {\bibfnamefont
  {C.}~\bibnamefont {Chin}}, \bibinfo {author} {\bibfnamefont {H.-C.}\
  \bibnamefont {N\"agerl}}, \ and\ \bibinfo {author} {\bibfnamefont
  {R.}~\bibnamefont {Grimm}},\ }\href@noop {} {\bibfield  {journal} {\bibinfo
  {journal} {Phys. Rev. A}\ }\textbf {\bibinfo {volume} {76}},\ \bibinfo
  {pages} {042514} (\bibinfo {year} {2007})}\BibitemShut {NoStop}%
\bibitem [{\citenamefont {Hodby}\ \emph {et~al.}(2005)\citenamefont {Hodby},
  \citenamefont {Thompson}, \citenamefont {Regal}, \citenamefont {Greiner},
  \citenamefont {Wilson}, \citenamefont {Jin}, \citenamefont {Cornell},\ and\
  \citenamefont {Wieman}}]{Hodby:2005}%
  \BibitemOpen
  \bibfield  {author} {\bibinfo {author} {\bibfnamefont {E.}~\bibnamefont
  {Hodby}}, \bibinfo {author} {\bibfnamefont {S.~T.}\ \bibnamefont {Thompson}},
  \bibinfo {author} {\bibfnamefont {C.~A.}\ \bibnamefont {Regal}}, \bibinfo
  {author} {\bibfnamefont {M.}~\bibnamefont {Greiner}}, \bibinfo {author}
  {\bibfnamefont {A.~C.}\ \bibnamefont {Wilson}}, \bibinfo {author}
  {\bibfnamefont {D.~S.}\ \bibnamefont {Jin}}, \bibinfo {author} {\bibfnamefont
  {E.~A.}\ \bibnamefont {Cornell}}, \ and\ \bibinfo {author} {\bibfnamefont
  {C.~E.}\ \bibnamefont {Wieman}},\ }\href@noop {} {\bibfield  {journal}
  {\bibinfo  {journal} {Phys. Rev. Lett.}\ }\textbf {\bibinfo {volume} {94}},\
  \bibinfo {pages} {120402} (\bibinfo {year} {2005})}\BibitemShut {NoStop}%
\bibitem [{\citenamefont {Z\"urn}\ \emph {et~al.}(2013)\citenamefont {Z\"urn},
  \citenamefont {Lompe}, \citenamefont {Wenz}, \citenamefont {Jochim},
  \citenamefont {Julienne},\ and\ \citenamefont
  {Hutson}}]{Zurn:Li2-binding:2013}%
  \BibitemOpen
  \bibfield  {author} {\bibinfo {author} {\bibfnamefont {G.}~\bibnamefont
  {Z\"urn}}, \bibinfo {author} {\bibfnamefont {T.}~\bibnamefont {Lompe}},
  \bibinfo {author} {\bibfnamefont {A.~N.}\ \bibnamefont {Wenz}}, \bibinfo
  {author} {\bibfnamefont {S.}~\bibnamefont {Jochim}}, \bibinfo {author}
  {\bibfnamefont {P.~S.}\ \bibnamefont {Julienne}}, \ and\ \bibinfo {author}
  {\bibfnamefont {J.~M.}\ \bibnamefont {Hutson}},\ }\href@noop {} {\bibfield
  {journal} {\bibinfo  {journal} {Phys. Rev. Lett.}\ }\textbf {\bibinfo
  {volume} {110}},\ \bibinfo {pages} {135301} (\bibinfo {year}
  {2013})}\BibitemShut {NoStop}%
\bibitem [{\citenamefont {Boisseau}\ \emph {et~al.}(1998)\citenamefont
  {Boisseau}, \citenamefont {Audouard},\ and\ \citenamefont
  {Vigu\'e}}]{Boisseau:1998}%
  \BibitemOpen
  \bibfield  {author} {\bibinfo {author} {\bibfnamefont {C.}~\bibnamefont
  {Boisseau}}, \bibinfo {author} {\bibfnamefont {E.}~\bibnamefont {Audouard}},
  \ and\ \bibinfo {author} {\bibfnamefont {J.}~\bibnamefont {Vigu\'e}},\
  }\href@noop {} {\bibfield  {journal} {\bibinfo  {journal} {Europhys. Lett.}\
  }\textbf {\bibinfo {volume} {41}},\ \bibinfo {pages} {349} (\bibinfo {year}
  {1998})}\BibitemShut {NoStop}%
\bibitem [{\citenamefont {Boisseau}\ \emph {et~al.}(2000)\citenamefont
  {Boisseau}, \citenamefont {Audouard}, \citenamefont {Vigu\'e},\ and\
  \citenamefont {Flambaum}}]{Boisseau:2000}%
  \BibitemOpen
  \bibfield  {author} {\bibinfo {author} {\bibfnamefont {C.}~\bibnamefont
  {Boisseau}}, \bibinfo {author} {\bibfnamefont {E.}~\bibnamefont {Audouard}},
  \bibinfo {author} {\bibfnamefont {J.}~\bibnamefont {Vigu\'e}}, \ and\
  \bibinfo {author} {\bibfnamefont {V.~V.}\ \bibnamefont {Flambaum}},\
  }\href@noop {} {\bibfield  {journal} {\bibinfo  {journal} {Eur. Phys. J. D}\
  }\textbf {\bibinfo {volume} {12}},\ \bibinfo {pages} {199} (\bibinfo {year}
  {2000})}\BibitemShut {NoStop}%
\end{thebibliography}%

\end{document}